\documentclass[12pt, draftclsnofoot, onecolumn]{IEEEtran}
\usepackage{epsfig,graphics,amssymb}
\usepackage{epsfig,graphics,amssymb}
\usepackage{amscd}
\usepackage{amsmath}
\usepackage{dsfont}

\usepackage{mathtools}
\usepackage{amscd}
\usepackage{amsmath}
\usepackage{enumerate}
\usepackage{theorem}
\usepackage{cite}
\usepackage[font={small}]{caption}
\usepackage{verbatim}
\usepackage{color}
\usepackage{paralist}

\theoremstyle{plain} \theorembodyfont{\itshape}
\theoremheaderfont{\scshape}

\theorembodyfont{\itshape} \theoremheaderfont{\scshape}
\newtheorem{lemma}{Lemma}

\theoremstyle{plain} \theorembodyfont{\itshape}
\theoremheaderfont{\scshape}

\newtheorem{assumption}{Assumption}

\newtheorem{proposition}{Proposition}

\usepackage[process=auto]{pstool}
\usepackage{tikz}

\newcommand{\matr}[1]{\mathlette{\boldmath}{#1}}
\newcommand{\mt}[1]{\mathbf{#1}}

\newcommand{\tr}{\mathrm{Tr}}

\newcommand{\argmin}{$\text{argmin}$}

\def\argmin{\mathop{\rm argmin}}

\def\expect{\mathop{\mbox{$\mathbb{E}$}}}

\newcommand{\e}{{\rm e}}

\newcommand{\Q}{{\rm Q}}

\newcommand{\define}{\stackrel{\triangle}{=}}

\newcommand{\R}{{\rm R}}
\newcommand{\dif}{{\rm d}}
\newcommand{\Dif}{{\rm D}}
\newcommand{\mathbbmss}[1]{\mathbb{#1}}

\newcommand{\T}{{\rm T}}

\title{Least Square Error Precoders for Massive~MIMO with Signal Constraints:\\ Fundamental Limits}
\author{\IEEEauthorblockN{Mohammad A. Sedaghat, Ali Bereyhi, Ralf R. M\"{u}ller, \emph{Senior Member, IEEE}}
\thanks{Mohammad A. Sedaghat, Ali Bereyhi and Ralf R. M\"uller are with the Institute for Digital Communications (IDC), Friedrich-Alexander Universit\"at Erlangen-N\"urnberg (e-mails: mohammad.sedaghat@fau.de, ali.bereyhi@fau.de, ralf.r.mueller@fau.de).}
}

\begin{document}

\maketitle


\begin{abstract}
This paper proposes the nonlinear Least Square Error (LSE) precoders for multiuser MIMO broadcast channels. The output signals of LSE Precoders are limited to be chosen from a predefined set which let these precoders address several constraints such as peak power limitation, constant envelope transmission and discrete constellations. We study the large-system performance of these precoders via the replica method from statistical physics, and derive a closed-form expression for the asymptotic distortion. Our results demonstrate that an LSE precoder with the output peak-to-average power ratio of $3~{\rm dB}$ can track the performance of the Regularized Zero Forcing (RZF) precoder closely. As the peak-to-average power ratio reduces to one, the constant envelope precoder is recovered. The investigations depict that the performance of the RZF precoder is achieved by the constant envelope precoder with $20\%$ of more transmit antennas. For $M$-PSK constellations, our analysis gives a lower-bound on the asymptotic distortion which is tight for moderate antenna-to-user ratios and deviates as the ratio grows. We improve this bound by deriving the replica solution under one-step of replica symmetry breaking. Our numerical investigations for this case show that the bound is tight for antenna-to-user ratios less than $5$.
\end{abstract}

\section{Introduction}
In massive Multiple-Input Multiple-Output (MIMO) systems, base stations employ a precoder in each coherence interval of the downlink channel to serve multiple users simultaneously \cite{rusek2013scaling}. The precoder is designed to minimize the mutual interference at the user terminals with respect to the available hardware constraints on the system. Consequently, the user terminals do not need to invoke complicated algorithms for detection, since the most of processing load is shifted from the user terminals to the transmit side. This fact has introduced the precoder as an essential element of massive MIMO systems whose user terminals are usually power-limited.

Several precoding schemes have been proposed so far which can be categorized into two classes of linear and nonlinear schemes. Linear schemes mainly consist of Match Filtering (MF), Zero Forcing (ZF) and Regularized Zero Forcing (RZF), where in practice each of them could be preferred regarding the desired tradeoff between the complexity and performance \cite{rusek2013scaling,peel2005vector}. As examples of nonlinear schemes, one can name Tomlinson-Harashima \cite{fischer2002space} and vector precoding \cite{peel2005vector}. The design of precoding schemes has been also investigated for cases in which the users' data symbols are taken from finite constellations, e.g., Phase Shift Keying (PSK) constellation~\cite{zeng2012linear}. 

Most of the precoders investigated in the literature are based on this assumption that the base station is able to transmit every possible signal. More precisely, the main body of work assumes that the signals at the output of the precoder can be chosen from the whole complex plane and are only limited in terms of average transmit power \cite{wiesel2008zero}. Nevertheless, this assumption does not hold in practice, since the precoded signals are transmitted via Radio Frequency (RF) chains and antennas which are restricted in several respects. For example, to improve the total power efficiency of base stations, nonlinear power amplifiers with low output back-off are desired. By using these amplifiers, the Peak-to-Average Power Ratio (PAPR) of the transmit signal should be kept low, in order to avoid nonlinear distortion on the signal. Therefore, for such a system, the precoder should be designed such that the output signals have low PAPRs. Another example is the recently proposed Load Modulated Single-RF (LMSRF) MIMO transmitter in which the signal on each antenna is taken from a predefined limited set \cite{sedaghat2016load,sedaghat:wsa2016}. In LMSRF transmitters, the number of switches in each load modulator determines the number of possible output constellation points; e.g., if every load modulator has only two switches; then, each transmit signal is chosen from four possible predefined signals. Here, one should note that this case is different from cases in which the users' data symbols have finite alphabet. In fact, in the case with finite alphabet data symbols, the precoded signal is not restricted and can take any value from the complex plane. 

Despite these examples, there are only few works in the literature in which the design of precoding schemes with respect to instantanous constraints on the transmit signal has been considered. In \cite{mohammed2013constant} and \cite{mohammed2013per}, the authors have proposed a nonlinear precoder whose precoded signals have a constant envelope on each transmit antenna. In another work in \cite{Globcom2014}, a new nonlinear precoder has been designed in order to limit the total instantaneous power at the transmitter of massive MIMO systems. In general, these lines of work as well as other set of constraints can be gathered in a unique framework. In fact, one can consider a class of precoders which select the precoded signals from a general set which includes all possible constraints on the transmit signals. These types of precoders have not been well studied so far, and to the best of our knowledge, there are no results in the literature which study the performance of this class of precoders analytically.
\subsection*{Contributions and Organization}
This paper investigates a general class of nonlinear precoders whose output signals are constrained to lie in some general set $\mathbb{X}$. For some given constraints on the output signals, the precoders find the transmit signals such that the total distortion at the user terminals is minimized. This class of precoders is therefore called the Least Square Error (LSE) precoders. We study the performance of these precoders in massive MIMO setups in which the system dimension grows large. 
The asymptotic distortion is derived analytically for a general set $\mathbb{X}$ by employing the replica method. Using the analytical results, we first consider the special form of LSE precoders which limits the PAPR of the precoded signal. For this case, the consistency of the asymptotic results with simulations is shown. Moreover, the constant envelope precoding scheme as a special case of the PAPR-limited LSE precoder is investigated, and a closed-form formula for its asymptotic distortion is derived which matches the results obtained via simulations. As another application of our results, we study a special form of LSE precoders in which the precoded symbols are limited to be taken from a PSK alphabet. For this case, it is shown that the asymptotic results given by the replica method bound the distortion from below. The bound is shown to be tight for small antenna-to-user ratio and start to deviate from the simulations as the antenna-to-user ratio increases. Our investigations depict that the analytical lower bound closely matches the simulations for the antenna-to-user ratio less than $5$. To investigate the tightness of the bound, we further derive a new lower bound using the union bound. The new bound is then shown to be outperformed by the lower bound evaluated via the replica method. 

The remaining parts of this paper is organized as follows: Section~\ref{sec:sys} introduces the LSE precoders for a massive MIMO downlink channel. In Section~\ref{sec:large}, the main results of the paper along with the large-system analysis are presented. Some special cases of the LSE precoders are then investigated explicitly in Section~\ref{sec:RS_special}. Section~\ref{sec:optimal} illustrates the rate maximization strategy for the LSE precoders, and Section~\ref{sec:num} presents the numerical results. Finally, the conclusion is given in Section~\ref{sec:conclusion}.

\subsection*{Notation}
We use bold lowercase and bold uppercase letters for vectors and matrices, respectively. $\mt{I}_K$ denoted the $K\times K$ identity matrix. The transposed and conjugate transposed of the matrix $\mt{H}$ are represented by $\mt{H}^{\sf T}$ and $\mt{H}^\dagger$, respectively. The set of real numbers is denoted by $\mathbb{R}$, and $\mathbb{C}$ identifies the complex plane. For the random vector $\matr{b}$, $F_{\matr{b}}(\cdot)$ is the cumulative distribution function (cdf). The Kronecker product is denoted by $\otimes$. The mutual information between the random variables $x$ and $y$ is represented by ${\rm I}(x;y)$, and the differential entropy of $x$ is denoted by ${\rm h}(x)$. $\Re$ and $\Im$ are used to identify the real and imaginary parts of a complex variable, respectively. $\expect$ represents the mathematical expectation, and the Gaussian averaging is abbreviated as
\begin{align}
\int (\cdot) \Dif z\define \frac1 \pi  \int (\cdot) \e^{-|z|^2}\dif z.
\end{align}
Moreover, we define ${\rm Vec}(\mt{A})$ to be the vector obtained by stacking the columns of $\mt{A}$.

\section{System Model and Problem Formulation}
\label{sec:sys}
We consider the general problem of precoding design for a single-cell massive MIMO system with $K$ single-antenna users and a base station which is equipped with $N$ antennas. 
Let $\matr{u}\in \mathbbmss{C}^K$ and $\mt{H}\in \mathbbmss{C}^{K\times N}$ denote the data vector of the users and the channel matrix, respectively. The precoded vector $\matr{v}$ is then evaluated from $\matr{u}$ and $\mt{H}$ such that $\matr{v}\in \mathbbmss{X}^N$ with $\mathbb{X}$ being a predefined set. Therefore, the received vector at the user terminals is written as
\begin{eqnarray}
\matr{y}=\mt{H} \hspace*{.3mm} \matr{v}+\matr{n},
\end{eqnarray}
where $\matr{y}=[y_1,\cdots,y_K]^\T$ with $y_k$ being the received signal at the user terminal $k$, and $\matr{n}$ is circularly symmetric zero-mean Gaussian noise with variance $\sigma_n^2$, i.e., $\matr{n}\sim\mathcal{CN} (\matr{0}, \sigma_n^2 \mt{I}_K)$.~Throughout the study, we assume that the data symbols of the users are independent and identically distributed (iid) Gaussian random variables, i.e., $\matr{u}\sim \mathcal{CN}(\matr{0},\sigma_u^2\mt{I}_K)$. It is moreover assumed that the channel is frequency-flat fading and perfectly known at the transmit side. The generalization to frequency-selective fading channels, as well as Orthogonal Frequency Division Multiplexing (OFDM) signals, is later presented in Appendix~\ref{appenfreuencyselective}. Therein, it is shown that the results derived for frequency-flat fading channels also hold for frequency-selective channels. 

The precoded vector $\matr{v}$ is evaluated via the nonlinear LSE precoder which for a given data vector $\matr{u}$ and channel matrix $\mt{H}$ reads
\begin{eqnarray}\label{precrule}
\matr{v}=\arg \min_{\matr{x} \in \mathbbmss{X}^N} \|\mt{H} \hspace*{.3mm} \matr{x}-\sqrt{\gamma} \hspace*{.5mm} \matr{u}  \|^2+\lambda \|\matr{x}\|^2 .
\end{eqnarray}
Here, $\gamma$ is a non-negative constant and $\lambda$ is a tuning parameter controlling the total transmit power\footnote{In fact, $\lambda$ is the Lagrange multiplier which takes the transmit power constraint into account.}. By setting $\mathbbmss{X}=\mathbbmss{C}$, the nonlinear LSE precoding scheme reduces to the linear scheme
\begin{align}\label{MMSEpr}
\matr{v}=\sqrt{\gamma} \hspace{.8mm}\mt{H}^\dagger \left(\mt{H} \mt{H}^\dagger +\lambda \mt{I}_K \right)^{-1}\matr{u}
\end{align} 
which is known as the RZF precoding scheme \cite{peel2005vector}. For the general set $\mathbbmss{X}$, however, the precoder is not of a simple form. The generality of $\mathbbmss{X}$ enables us to model various signal constraints in MIMO transmitters. Some examples of these constraints are as follows.
\begin{itemize}
\item As the first example, one can model peak power constraints on each antenna by defining
\begin{equation}
\mathbbmss{X}=\left\{x=r\e^{j\theta} : \hspace*{.5mm} \theta\in[0,2\pi] \hspace*{2mm} \text{and} \hspace*{2mm} 0\leq r\leq \sqrt{P}\right\},
\end{equation}
where $P$ is the maximum power at each antenna. Noting that the average transmit power is constrained by the tuning factor $\lambda$, the output PAPR is also limited in this case.
\item Precoding over with finite discrete constellations for LMSRF MIMO transmitters \cite{sedaghat2014novel,sedaghat2016load} can be realized by setting $\mathbb{X}$ to be the finite set of modulators' states.

\item Per-antenna constant envelope precoding \cite{mohammed2013per} is another example where
\begin{eqnarray}
\vert v_i\vert ^2=P \quad \forall i \in\{1,\cdots,N\}.
\end{eqnarray}
In this case, the precoded symbols have a constant amplitude. Consequently, depending on the pulse shaping filter, the PAPR of the output signal can be approximately reduced to $3$ dB, and thus, highly efficient nonlinear power amplifiers can be utilized. 
\end{itemize}

These examples, as well as other constraints on transmit signals, can be addressed via the class of LSE precoders. 
%
Considering the employment of this precoding scheme in massive MIMO setups, we are interested in studying the performance of the precoders in the large-system limit\footnote{By the large-system limit, we mean $K$ and $N$ grow large while the ratio $\alpha=N/K$ is kept fixed.}. We quantify the performance by defining the asymptotic distortion at user terminals as the measure. For the precoded vector $\matr{v}$ and its corresponding data vector $\matr{u}$ and channel $\mt{H}$, the asymptotic distortion per user is given by
\begin{eqnarray}\label{orgopt}
D =\lim_{K\uparrow \infty} \frac{1}{K} \expect\|\mt{H} \hspace*{.5mm} \matr{v}-\sqrt{\gamma} \hspace*{.5mm} \matr{u}  \|^2
\end{eqnarray}
when the antenna-to-user ratio\footnote{One may call this ratio the inverse load factor, since $K/N$ is usually referred to as the load factor in massive MIMO systems.}, defined as $\alpha\define {N}/{K}$, is kept fixed.
The asymptotic distortion can be used to derive a lower bound on the ergodic achievable rate of the users in the downlink channel when the LSE precoding scheme is employed. To state the bound, let $R_k$ be the ergodic achievable rate of user $k$, and $z_k\left(\mt{H},\matr{u}\right)$ be the interference at this user terminal. The received signal in this case can be written as
\begin{eqnarray}
y_k=\sqrt{\gamma} \hspace*{.5mm} u_k+z_{k}(\mt{H},\matr{u})+n_k,
\end{eqnarray}
where $u_k$ and $n_k$ denote the $k$th entry of $\matr{u}$ and $\matr{n}$, respectively. The average ergodic rate $\tilde{R}$ is then defined as
\begin{align}
\tilde{R}&\define \frac{1}{K}\sum_{k=1}^K  R_k =\frac{1}{K}\sum_{k=1}^K {\rm I} (u_k;y_k).
\end{align}
With some lines of derivations, it is shown that the average ergodic rate is bound from below as stated in the following lemma.
\begin{lemma}
\label{khodlemmalowerrate}
As $K\uparrow \infty$, the average ergodic rate reads
\begin{eqnarray}
\tilde{R} \geq \log\left(\frac{\gamma\sigma_u^2}{\sigma_n^2+D}\right).
\end{eqnarray}
\end{lemma}
\textbf{Proof.} The proof is given in Appendix~\ref{lemmalowerrate}.\\

Determining the asymptotic distortion is not a trivial task. In fact, the classical analytic tools fail to analyze the optimization problem in \eqref{precrule} for a lot of choices of $\mathbb{X}$. We therefore invoke the replica method developed in statistical mechanics to study the performance of precoders. The replica method enables us to evaluate the asymptotic distortion defined in \eqref{orgopt} without finding the explicit solution of the optimization problem in \eqref{precrule}. Although the exact precoded vector cannot be found through this replica-based analysis, it is still important to find an estimate of the performance, as it can give a reference measure to investigate the efficiency of available algorithms used in practice. In the next section, we employ the replica method and derive the asymptotic distortion of the LSE precoder for a general set $\mathbb{X}$. 

\section{Large-system Analysis of LSE Precoders}
\label{sec:large}
The main goal of this section is to determine the asymptotic distortion $D$ for a general set $\mathbb{X}$. We start the analysis by stating some simple definitions. Let the matrix $\mt{R}$ be the Gramian of $\mt{H}$ which is written as $\mt{R}\define \mt{H}^\dagger \mt{H}$. The asymptotic parameter $\tilde{D}$ is defined as
\begin{eqnarray}
&&\tilde{D}\define\lim_{K\uparrow \infty}\frac{1}{K}\expect\min_{\matr{x} \in \mathbbmss{X}^N} \|\mt{H} \hspace*{.5mm} \matr{x}-\sqrt{\gamma} \hspace*{.5mm} \matr{u}  \|^2 + \lambda \| \matr{x}\|^2. 
\end{eqnarray}
Later in this section, we show that $D$ can be calculated easily from $\tilde{D}$. We further define 
\begin{eqnarray}
g(\matr{x})\define\matr{x}^\dagger \mt{R} \ \matr{x}-2\sqrt{\gamma} \hspace*{.5mm} \Re\left\{\matr{x}^\dagger \mt{H}^\dagger \matr{u}\right\}+ \lambda \matr{x}^\dagger \matr{x}.
\end{eqnarray}
Noting that $\lim\limits_{K\uparrow \infty} {\matr{u}^\dagger \matr{u}}/K=\sigma_u^2$, one can show that $\tilde{D}$ reads
\begin{align}\label{optinitialrep}
\tilde{D}=\gamma\sigma_u^2+\lim_{K\uparrow \infty}\frac{1}{K}\expect\min\limits_{\matr{x} \in \mathbbmss{X}^N} g(\matr{x}).
\end{align}
To determine the parameter $\tilde{D}$, we employ the following lemma which is concluded using some standard large deviations arguments.
\begin{lemma}\label{varahlemma}
The minimization in the right hand side of \eqref{optinitialrep} can be written as
\begin{align}\label{Farhadkhan}
&\min\limits_{\matr{x}\in \mathbbmss{X}^N}{g} (\matr{x})
=-\lim\limits_{\beta\uparrow \infty}\frac{1}{\beta}
 \log\int\limits_{\matr{x} \in \mathbbmss{X}^N}\e^{-\beta{g} (\matr{x})} \dif \matr{x}.
\end{align}
\end{lemma}
\textbf{Proof.} This argument is a special case of Varadhan's theorem in large deviations theory and is obtained by letting $\epsilon=1/\beta$, $\phi(x)=0$ and $\{\mu_{\epsilon}\}$ equal to a family of non-degenerate Gaussian measures with the rate function $g(\matr{x})$ in \cite[ Theorem~4.3.1]{dembo1998large}.\\


The large deviations argument in \eqref{Farhadkhan} implies that as $\beta\uparrow \infty$ only one of the vectors in the set $\mathbb{X}^N$ dominates the integral\footnote{In fact, \eqref{Farhadkhan} can be considered as the generalization of the simple equality $\min(x_1,x_2)=-\lim\limits_{\beta\uparrow \infty} 1/\beta \log\left(\e^{-\beta x_1} +\e^{-\beta x_2} \right).$}. Using Lemma~\ref{varahlemma}, \eqref{optinitialrep} is rewritten as
\begin{align}\label{dhat2}
\tilde{D}=\gamma\sigma_u^2-\lim_{\beta,K\uparrow \infty}\frac{1}{K\beta}\expect   \log\int\limits_{\matr{x} \in \mathbbmss{X}^N}\e^{-\beta{g} (\matr{x})} \dif \matr{x}.
\end{align}
From \eqref{dhat2}, one can observe that the calculation of $\tilde{D}$ needs a logarithmic expectation to be determined. This latter task is not trivial for a general set $\mathbbmss{X}$. We therefore employ well-known Riesz equality which is used as the start point of every study based on the replica method \cite{mezard2009information,merhav2010statistical}. The equality implies that for a given non-negative random variable $t$, we have
\begin{equation}\label{replicacont}
\expect \log(t) =\lim_{n\downarrow 0}\frac{\partial}{\partial n} \log \expect t^n.
\end{equation}
By applying the equality in \eqref{replicacont} to \eqref{dhat2}, $\tilde{D}$ is determined as
\begin{subequations}
\begin{align}\label{defEn}
\tilde{D}&= \gamma\sigma_u^2 -\lim_{K,\beta\uparrow \infty}\frac{1}{\beta K}\lim_{n\downarrow 0}\frac{\partial }{\partial n} 
\log \expect \left[\int\limits_{~\matr{x} \in \mathbbmss{X}^N}\e^{-\beta g(\matr{x})  } \dif\matr{x}\right]^n \\
&= \gamma\sigma_u^2 -\lim_{\beta\uparrow \infty}\frac{1}{\beta}\lim_{n\downarrow 0}\frac{\partial }{\partial n} \Xi_n,
\end{align}
\end{subequations}
where $\Xi_n$ is defined as 
\begin{eqnarray}
\Xi_n\define
\lim_{K\uparrow \infty} \frac{1}{K} \log \expect\left[\int\limits_{~\matr{x} \in \mathbbmss{X}^N}\e^{-\beta g(\matr{x})  }\dif \matr{x}\right]^n. 
\end{eqnarray}
By noting that $n$ in \eqref{replicacont} is tending to zero on the real axis, one needs to determine $\Xi_n$ for real values of $n$, in order to find $\tilde{D}$ from \eqref{defEn}. The replica method suggests us to find $\Xi_n$ by considering the replica continuity assumption.
\begin{assumption}[Replica Continuity]
Assume that the function $\Xi_n$ analytically continues from the set of non-negative integers $\mathbb{Z}^+$ onto the real axis $\mathbb{R}$. This means that one can calculate $\Xi_n$ for integer values of $n$, and assume that the expression also holds for real values of $n$.\footnote{In fact, the assumption does not need to hold on whole $\mathbb{R}$, and is sufficient to hold within a right neighborhood of $n =0$.}
\end{assumption}

The replica continuity assumption has not been yet rigorously justified for a general case \cite{mezard2009information}. However, using some alternative mathematical tools, many results derived via th replica method have been shown to be exact \cite{mezard2009information, talagrand2006parisi}. It is therefore strongly believed in the literature that replica continuity holds at least for the forms objective functions which we have considered in this study. Using numerical investigations, we later show that the results derived under this assumption are consistent in several particular cases with simulations. 

Considering replica continuity to hold, $n$ is assumed to be an integer. Thus, $\Xi_n$ reads
\begin{align}\label{Xin}
&\Xi_n
=\lim_{K\uparrow \infty} \frac{1}{K} \log  
 \int\limits_{\{\matr{x}_a \}} \expect \e^{-\beta  \sum\limits_{a=1}^n g(\matr{x}_a)    } \dif\matr{x}_1\cdots \dif\matr{x}_n
\end{align}
where the notation $\{\matr{x}_a \}$ denotes the vector of replicas $\{\matr{x}_1,\ldots,\matr{x}_n\} \in \mathbbmss{X}^N\times \ldots \times \mathbbmss{X}^N$. In fact, this is the main reason that the method is called the replica method. Using the independency of $\matr{u}$ and $\mt{H}$, the expectations over $\matr{u}$ and $\mt{H}$ separate. Thus, by taking the expectation over $\matr{u}$ with Gaussian iid entries, \eqref{Xin} can be written as
\begin{align}
&\Xi_n 
 =\lim_{K\uparrow \infty} \frac{1}{K} \log   \int\limits_{\{\matr{x}_a \}} \expect_\mt{H} \e^{-\beta\sum\limits_{a=1}^{n}\left[\matr{x}_a^\dagger \mt{R} \hspace{.5mm} \matr{x}_a+ \lambda \matr{x}_a^\dagger \matr{x}_a\right] +\beta^2\gamma\sigma_u^2 \|\sum\limits_{a=1}^n \mt{H} \matr{x}_a\|^2}\dif\matr{x}_1\cdots \dif\matr{x}_n,
\end{align}
where $\expect\limits_{\mt{H}}$ denotes the expectation with respect to $\mt{H}$.
We further define the matrix $\mt{V}$ to be
\begin{eqnarray}\label{Veqn}
\mt{V}\define\frac 1N \left[ \matr{x}_1,\cdots,\matr{x}_n \right] \mt{\Gamma} \left[ \matr{x}_1,\cdots,\matr{x}_n \right]^\dagger
\end{eqnarray} 
where $\mt\Gamma$ is an $n\times n$ matrix whose entry $(a,b)$ reads
\begin{eqnarray}
\zeta_{ab}\define -\beta\gamma\sigma_u^2 +\delta_{a,b}
\end{eqnarray}
with $\delta_{a,b}=1$ for $a=b$ and being zero elsewhere. Consequently, $\Xi_n$ is written as 
\begin{align}\label{beforeHarish}
&\Xi_n =\lim_{K\uparrow \infty}\frac{1}{K} \log  \int\limits_{\{\matr{x}_a \}}\e^{-\beta \lambda \sum_{a=1}^n\matr{x}_a^\dagger \matr{x}_a}\expect_\mt{H} \e^{-\beta N \tr\left( \mt{R}  \mt{V} \right)} \dif\matr{x}_1\cdots \dif\matr{x}_n.
\end{align}
As the next step, we need to calculate the expectation in \eqref{beforeHarish} with respect to $\mt H$. To this end, we first need to define the Stieltjes transform and the $\rm R$-transform of a given distribution. Suppose that the empirical eigenvalue distribution of the matrix $\mt{R}$ converges to a deterministic distribution, and denote the corresponding cdf with $F_{\mt{R}}(\lambda)$. The Stieltjes transform of the distribution $F_{\mt{R}}(\lambda)$ is defined as ${\mathrm G}_{\mt{R}}(s)=\expect (\lambda-s)^{-1}$ where the expectation is taken with respect to $F_{\mt{R}}(\lambda)$ \cite{tulino2004random}. The corresponding $\mathrm{R}$-transform is then defined as \cite{voiculescu1986addition}
\begin{align}\label{Rstilet}
\R_{\mt{R}}(w)=\mathrm G_{\mt{R}}^{-1}(w)-w^{-1}
\end{align}
where $\mathrm G_{\mt{R}}^{-1}(w)$ denotes the inverse with respect to composition.

The expectation in \eqref{beforeHarish} is in fact a spherical integral which is known as the Harish-Chandra-Itzykson-Zuber integral in physics and mathematics literature. In the large-system limit,~the~integral has been calculated in \cite{guionnet:05}. Using the results given in \cite{guionnet:05} the expectation is calculated~as
\begin{eqnarray}
\expect_\mt{H} \e^{-\beta N \tr\left( \mt{R}  \mt V\right)}=\e^{-N \sum\limits_{i=1}^N \int\limits_0^{\beta \tilde{\lambda}_i}\R_{\mt{R}}(-w)\dif w + o(N)} \label{eq:Harish2}
\end{eqnarray}
where $\tilde{\lambda}_1,\cdots, \tilde{\lambda}_N$ are the eigenvalues of $\mt{V}$ and $\lim_{N\uparrow\infty}o(N)/N=0$.~Int- erested readers are referred to \cite[Appendix~F]{bereyhi2016statistical}, for more detailed discussions on the asymptotics of spherical integrals. Considering the right hand side of \eqref{eq:Harish2}, one observes that the matrix $\mt{V}$ has only $n$ nonzero eigenvalues which are equal to the eigenvalues of the matrix\footnote{This can be easily shown using the fact that the nonzero eigenvalues of the two matrices $\mt{AB}$ and $\mt{BA}$ are the same.} 
\begin{eqnarray}
\mt{G}=\frac{1}{N} \ \mt\Gamma  \left[\matr{x}_1,\cdots,\matr{x}_n\right]^\dagger\left[\matr{x}_1,\cdots,\matr{x}_n\right]. \label{matG}
\end{eqnarray}
Therefore, by defining $\lambda_1,\cdots, \lambda_n$ to denote the eigenvalues of $\mt{G}$, \eqref{eq:Harish2} reduces to
\begin{eqnarray}
\expect_\mt{H} \e^{-\beta N \tr\left( \mt{R}  \mt V\right)}=\e^{-N \sum\limits_{i=1}^n \int\limits_0^{\beta \lambda_i}\R_{\mt{R}}(-w)\dif w +o(N)}. \label{eq:harishend}
\end{eqnarray}
By substituting \eqref{eq:harishend} in \eqref{beforeHarish}, one can observe that the derivation of $\Xi_n$ further needs to calculate an integral over the $Nn$-dimensional space. We determine the integral by splitting the space into subshells, such that every two replicas within each sunshell have a fixed correlation \cite{mueller:08}. We therefore define the subshell $\mathcal{S}(\mt{Q})$ to be
\begin{eqnarray}
\mathcal{S}(\mt{Q})\define \{\matr{x}_1,\cdots,\matr{x}_n|\matr{x}_a^\dagger \matr{x}_b=NQ_{ab}\},
\end{eqnarray}
in which $Q_{ab}$ is the entry $(a,b)$ of the replicas' correlation matrix $\mt Q$ defined as
\begin{eqnarray}
\mt{Q}=\frac{1}{N} \ [\matr{x}_1,\cdots,\matr{x}_n]^\dagger[\matr{x}_1,\cdots,\matr{x}_n].
\end{eqnarray}
Consequently, one can change the integration variable from $[\matr{x}_1,\cdots,\matr{x}_n]$ to $\mt Q$ and write \eqref{beforeHarish} as 
\begin{align} \label{replIQ}
&\Xi_n 
= \lim_{K\uparrow \infty}\frac{1}{K} \log   \int \e^{N\mathcal{I}(\mt{Q})} \e^{-N\mathcal{G}(\mt{Q})}\mathcal{D}\mt{Q},
\end{align}
where the function $\mathcal{G}(\mt{Q})$ is defined as
\begin{eqnarray}
\mathcal{G}(\mt{Q})\define\beta \lambda \sum_{a=1}^n\frac{\matr{x}_a^\dagger \matr{x}_a}{N} +\sum\limits_{a=1}^n \int\limits_0^{\beta \lambda_a}\R_\mt{R}(-w)\dif w ,
\end{eqnarray}
$\mathcal{D} \mt Q$ reads
\begin{equation}
{\cal D}\mt Q \define \prod\limits_{a=1}^{n}\dif Q_{aa} \prod \limits_{b=a+1}^n {\rm d}{\Re} \left\lbrace Q_{ab} \right\rbrace {\rm d}{\Im} \left\lbrace Q_{ab} \right\rbrace 
\end{equation}
and $\e^{N\mathcal{I}(\mt{Q})}$ is the Jacobian term appears by the change of integration variable. By taking the same approach as in \cite{mueller:08}, the Jacobian term is calculated as
\begin{align}
\hspace*{-2mm}{\rm e}^{N\mathcal{I}(\matr Q)} \hspace*{-.3mm} = \hspace*{-2mm} \int\limits_{\{\matr{x}_a\}} \hspace*{-1mm}
 \prod\limits_{a=1}^n \hspace*{-.7mm} \delta\left(\matr{x}_a^\dagger\matr{x}-NQ_{a,a} \right) \hspace*{-2mm} \prod_{b=a+1}^n \hspace*{-2mm} \delta\left( \Re\left\lbrace {\matr{x}}_a^\dagger {\matr{x}}_b-NQ_{ab}\right\rbrace  \right) 
\delta\left( \Im\left\lbrace{\matr{x}}_a^\dagger {\matr{x}}_b-N Q_{ab}\right\rbrace  \right)\ \hspace*{-2mm} \prod_{a=1}^{n} \hspace*{-.7mm} {\rm d}\matr{x}_a.
\end{align}
In order to determine the Jacobian term, we define the matrix $\tilde{\mt{Q}}$ whose diagonal entries are $\tilde{Q}_{aa}=\tilde{Q}_{aa}^I$ and off-diagonal entries are $\tilde{Q}_{ab}=\frac12(\tilde{Q}_{ab}^I - j \tilde{Q}_{ab}^Q)$ and $\tilde{Q}_{ba}=\frac12(\tilde{Q}_{ab}^I + j \tilde{Q}_{ab}^Q)$ for some complex $\tilde{Q}_{ab}^I$ and $\tilde{Q}_{ab}^Q$. Following the lines of derivation in \cite[eq.(52)-(58)]{mueller:08} and defining $\mathcal{J}\define(t-j\infty;t+j\infty)$ for some $t\in \mathbbmss{R}$, we obtain
\begin{eqnarray}
\e^{N\mathcal{I}(\mt{Q})}=\int\limits_{\mathcal{J}^{n^2}}\e^{-N\tr[\tilde{\mt{Q}} \mt{Q}]+N\log \mathcal{M } (\tilde{\mt{Q}})} \tilde{\mathcal{D}}\tilde{\mt{Q}} \label{eq:jacobi}
\end{eqnarray}
where the function $\mathcal{M}(\tilde{\mt{Q}})$ is defined to be
\begin{eqnarray} \label{tarifM}
\mathcal{M}(\tilde{\mt{Q}})\define\sum\limits_{\{x_a\}} \e^{\hspace*{0.5mm} \sum\limits_{b=1}^n\sum\limits_{a=1}^n x_a^*x_b\tilde{Q}_{ab}}
\end{eqnarray}
for $x_a\in\mathbb{X}$ and $\{x_a\}$ denoting $\{x_1,x_2\cdots,x_n\}\in\mathbb{X}\times\mathbb{X}\times \cdots \times \mathbb{X}$. Moreover, $\tilde{\mathcal D}\tilde{\mt Q}$ reads \cite{mueller:08}
\begin{equation}
\tilde{\cal D} \tilde{\mt Q} \define \prod\limits_{a=1}^{n} \frac{\dif Q_{aa}^I}{2\pi j} \prod \limits_{b=a+1}^n \frac{{\rm d} Q_{ab}^I  {\rm d}Q_{ab}^Q}{(2\pi j)^2}.
\end{equation}
By replacing \eqref{eq:jacobi} in \eqref{replIQ}, we can find $\Xi_n$ in terms of the large-system limit of an integral. Using Lemma~\ref{varahlemma}, the integration in \eqref{replIQ} is dominated by the integrand at the saddle-point as $N$ and $K$ grow large. For general form of $\mt{Q}$ and $\tilde{\mt{Q}}$, calculating the saddle-point of the integrand is a complicated task. The replica method therefore suggests us to assume a predefined structure on $\mt{Q}$ and $\tilde{\mt{Q}}$ and search for the saddle-point within these classes of matrices. As it is well-known from the literature, Replica Symmetry (RS) proposes the most primary structure. The structure, however, may not result in the true saddle point. In that case, we need to employ Replica Symmetry Breaking scheme to widen recursively the set over which we search for the saddle-point. In the sequel, we start derive the asymptotic distortion $D$ by considering the RS structures as well as RSB with one step of recursion.


%
%
%
\subsection{Asymptotic Distortion under Replica Symmetry}
\label{sec:RS}
The most primary structure is imposed by the RS assumption. In the RS assumption, it is postulated that the solutions of $\mt{Q}$ and $\tilde{\mt{Q}}$ which dominate the integral in \eqref{replIQ} are invariant to permutation of the replica indices. This means that $Q_{aa}$ is the same for all indices $a\in\left\lbrace 1, \ldots, n \right\rbrace$, and also $Q_{ab}\neq Q_{aa}$ are the same for all $a\neq b$. This is a very simple structure which has given amazingly exact solutions in many cases \cite{mezard2009information}. The intuition behind the proposal of such a structure comes from physical interpretation of the replica analyses \cite{mezard2009information}. Following the notations in \cite{mueller:08} under the RS assumption, we set $Q_{ab}=q$ and $\tilde{Q}_{ab}=\beta^2 f^2$ for all $a\neq b$ and $Q_{aa}=q+\chi/\beta$ and $\tilde{Q}_{aa}=\beta^2f^2-\beta e$ for $a\in\left\lbrace 1, \ldots, n \right\rbrace$.
Here, $q$, $\chi$, $f$ and $e$ are some non-negative real variables which parameterize the correlation matrices and need to be calculated at the saddle-point. By substituting the RS structure in \eqref{replIQ}, $\Xi_n$ can be analytically calculated, and consequently, $\tilde{D}$ is determined. 
The asymptotic distortion under the RS assumption is stated in Proposition~\ref{prop1}.

\begin{proposition}\label{prop1}
\normalfont
Under the RS assumption, the asymptotic distortion is determined by
\begin{eqnarray}
D_{\rm RS}=\gamma\sigma_u^2+\alpha \hspace{.5mm} \frac{\partial}{\partial \chi}\left[ (q-\chi\gamma\sigma^2_u)\chi \R_{\mt{R}}(-\chi)\right].
\end{eqnarray}
The scalars $q$ and $\chi$ are solutions to the following set of fixed-point equations
\begin{subequations}
\begin{align}\label{RScoupledEqs1}
&\chi=\frac 1f \Re \int_{\mathbbmss{C}} \argmin\limits_{x\in \mathbbmss{X}} \left|z-\frac{\R_{\mt{R}}(-\chi)+\lambda }{f}x  \right|z^* \Dif z   \\
&q=\int_{\mathbbmss{C}} \left|\argmin\limits_{x\in \mathbbmss{X}} \left|z-\frac{\R_{\mt{R}}(-\chi)+\lambda}{f} x \right| \right|^2 \Dif z\label{RScoupledEqs2}
\end{align}
\end{subequations}
where $f$ is determined in terms of $\chi$ and $q$ as
\begin{eqnarray}
f\define \sqrt{(q-\chi \gamma\sigma_u^2)\R^\prime_\mt{R}(-\chi)+\gamma\sigma^2_u \R_{\mt{R}}(-\chi)}.
\end{eqnarray}
\end{proposition}
\textbf{Proof.} The proof is given in Appendix~\ref{app1}.\\

Considering the distortion derived under the RS assumption, one needs to investigate whether $D_{\rm RS}$ returns the exact value of $D$. To answer this question, we need to give a brief overview on the possible reference points and consistency tests in the literature. Throughout the literature of multiuser communications, the RS assumption has been proven to give the exact solution in many problems using some alternative mathematical methods. The well-known example is the pioneering work by Tanaka which employed the replica method under the RS assumption to calculate the input-output mutual information in a Code Division Multiple-Access (CDMA) multiuser system with iid binary inputs \cite{tanaka:02}. Few years later, Tanaka's formula could be confirmed by using some complicated mathematical methods \cite{montanari2006analysis}. Nevertheless, such alternative methods are not always available, and thus, it is beneficial to calculate the solution suggested by the replica method. To find the scope of setups in which the RS solution returns the exact distortion, we invoke the strong belief in the literature which conjectures that the solutions of convex optimization problems are perfectly determined by the replica method under the RS assumption \cite{moustakas2007outage,zaidel2012vector}. Our numerical investigations in the next sections confirms the validity of this conjecture in our case. There are, however, several examples in the literature in which the replica method under the RS assumption fails to give exact solution\footnote{So far, these problems have been non-convex and NP-hard.} \cite{mezard2009information,zaidel2012vector}. For such cases, Parisi in \cite{parisi1980sequence} introduced the RSB scheme to recursively extends the saddle-point's search set. The scheme was later proven to give the exact solution for the Sherrinton-Kirkparick (SK) model \cite{talagrand2006parisi}. For this model, Guerra showed in \cite{guerra2001sum} that the solution under RS is always a lower bound. In the following, we derive the asymptotic distortion under RSB with one recursion. We later show by numerical simulations that the asymptotic distortion for our setup for the case with PSK output constellation needs RSB to be considered. Moreover, similar to Guerra's result, we observe that the $D_{\rm RS}$ in this case is a lower bound on $D$.
%
%
%
%
\subsection{Asymptotic Distortion under Replica Symmetry Breaking}
\label{sec:RSB}
The RSB scheme starts from the RS structure and extends the set of possible choices for $\mt Q$ and $\tilde{\mt Q}$ at the saddle-point recursively. The structure obtained after $r$ steps of recursion is referred to as $r$-RSB structure (or assumption). It has been shown in the literature that the replica method under the assumption of full-RSB, which means $r$-RSB when $r\uparrow \infty$, gives the exact solution for the SK model \cite{talagrand2006parisi}; see \cite[Section 8.2]{mezard2009information} for the SK model. In \cite{guerra2003broken}, Guerra generalized the result in \cite{guerra2001sum} for RSB and showed that the replica method gives a lower bound under $r$-RSB assumption with any finite $r$ for the SK model. This result was later generalized for a more general model \cite{franz2003replica}. Our numerical investigations agree with Guerra's argument, as we observe that for the $M$-PSK constellation the distortion, which is evaluated numerically via simulations in the large-system limit, is bounded by the RS and $1$-RSB solutions from below. The lower-bound given via $1$-RSB is a tighter bound compared to that of RS.

Following the notations of \cite{zaidel2012vector}, for RSB scheme with one recursion, $\mt Q$ and $\tilde{\mt Q}$ are 
\begin{subequations}
\begin{align} 
\mt{Q}&=q_1 \matr{1}_{n}+p_1\mt{I}_{\frac{n\beta}{\mu_1}} \otimes \matr{1}_{\frac{\mu_1}{\beta}}+\frac{\chi_1}{\beta} \mt{I}_{n}, \label{RSBstr1} \\
\tilde{\mt{Q}}&=\beta^2 f_1^2 \matr{1}_{n} + \beta^2 g_1^2\mt{I}_{\frac{n\beta}{\mu_1}} \otimes \matr{1}_{\frac{\mu_1}{\beta}} -\beta e_1 \mt{I}_n,\label{RSBstr2}
\end{align}
\end{subequations}
where $q_1$, $p_1$, $\chi_1$, $\mu_1$, $f_1$, $g_1$ and $e_1$ are non-negative real scalars and $\matr{1}_{n}$ is an $n\times n$ all-ones matrix. A schematic illustration of $1$-RSB structure is given in \cite[Fig. 8.4]{mezard2009information}. Prposition~\ref{prop2} states the result for the asymptotic distortion when the $1$-RSB structure is employed.

\begin{proposition}\label{prop2}
\normalfont
Under the 1-RSB assumption, the asymptotic distortion is obtained as
 \begin{align}
\hspace*{-2mm}D_{\rm RSB} \hspace*{-1mm} =\hspace*{-1mm} \gamma\sigma_u^2 \hspace*{-1mm} - \hspace*{-1mm} \frac{\alpha\chi_1}{\mu_1}
\R_{\mt{R}}(-\chi_1) \hspace*{-1mm}
+ \hspace*{-1mm} \alpha\left[ q_1+\frac{\eta_1}{\mu_1}-2\gamma\sigma_u^2\eta_1\right]\R_{\mt{R}}(-\eta_1)  \hspace*{-1mm}
-  \hspace*{-1mm} \alpha\eta_1\left[q_1 \hspace*{-.5mm} - \hspace*{-.5mm} \gamma\sigma_u^2\eta_1\right]\R_{\mt{R}}^{\prime}(-\eta_1).
\end{align}
The set of scalars $\{q_1,p_1,\chi_1,\mu_1\}$ is calculated through the fixed-point equations
\begin{subequations}
\begin{eqnarray}
\eta_1&=&\frac{1}{f_1}\int \int\Re\left\{z^*\argmin\limits_{x\in \mathbbmss{X}}|f_1z+g_1y-e_1x |   \right\}\tilde{\mathcal{Y}}(y,z)\Dif z \Dif y, \\
q_1+p_1&=&\int \int\left|\argmin\limits_{x\in \mathbbmss{X}}|f_1z+g_1y-e_1x |   \right|^2\tilde{\mathcal{Y}}(y,z)\Dif z \Dif y,  \\
\eta_1+\mu_1q_1&=& \frac{1}{g_1}\int \int\Re\left\{y^*\argmin\limits_{x\in \mathbbmss{X}}|f_1z+g_1y-e_1x |   \right\}\tilde{\mathcal{Y}}(y,z)\Dif z \Dif y,
\end{eqnarray}
and
\begin{align}
\int_{\chi_1}^{\eta_1} \R_{\mt{R}}(-w)\dif w=&\int \log \int \mathcal{Y}(y,z)\Dif y \Dif z+(\mu_1q_1+2\eta_1-2\mu_1 \eta_1 \gamma\sigma_u^2-2\chi_1\mu_1\gamma\sigma_u^2)\R_{\mt{R}}(-\eta_1)\nonumber \\
&-2\chi_1\R_{\mt{R}}(-\chi_1) -2\mu_1\eta_1(q_1-\gamma\sigma_u^2\eta_1)	\R_{\mt{R}}^{\prime}(-\eta_1) 
+ \lambda\mu_1(p_1+q_1),
\end{align}
\end{subequations}
where $\eta_1=\chi_1+\mu_1p_1 $, the function $\mathcal{Y}(y,z)$ is given by
\begin{eqnarray}
\mathcal{Y}(y,z)=\e^{-\mu_1 \min\limits_{x\in \mathbbmss{X}}e_1 |x|^2-2\Re\{x(f_1z^*+g_1y^*)\}}
\end{eqnarray}
and $\tilde{\mathcal{Y}}(y,z)$ is defined as
\begin{eqnarray}
\tilde{\mathcal{Y}}(y,z)=\frac{\mathcal{Y}(y,z)}{\int_{\mathbbmss{C}}  \mathcal{Y}(\tilde{y},z)\Dif \tilde{y}}.
\end{eqnarray}
Moreover, the parameters $e_1$, $f_1$ and $g_1$ are determined as
\begin{subequations}
\begin{align}
&e_1=\R_{\mt{R}}(-\chi_1)+\lambda \\
&f_1=\sqrt{\gamma\sigma_u^2\R_{\mt{R}}(-\eta_1) + (q_1-\gamma\sigma_u^2\eta_1)\R_{\mt{R}}^{\prime}(-\eta_1)} \\
&g_1=\sqrt{\frac{\R_{\mt{R}}(-\chi_1)-\R_{\mt{R}}(-\eta_1)}{\mu_1}}.
\end{align}
\end{subequations}
\end{proposition}
\textbf{Proof.} The proof is given in Appendix~\ref{rsbproof}.\\


The fixed-point equations in the both RS and 1-RSB cases may have multiple~solutions.~However, only one of them determines the valid saddle-point of \eqref{replIQ}. In this case,  we note that due to Varadhan's theorem the saddle-point solution is the one which minimizes the asymptotic distortion. In fact, among the possible solutions, it is the global optimum which dominates the integral in the large-system limit and the effect of the other solutions disappears and $N,K\uparrow\infty$.


A common test in statistical mechanics for validating the prediction of the replica method for systems with discrete states is to check the entropy of the corresponding thermodynamic system \cite{zaidel2012vector}. In this test, the so-called ``zero-temperature entropy'' is checked to see weather it vanishes. The intuition behind this test comes from the fact that the entropy of the thermodynamic system which corresponds to the LSE precoders should tend to zero as the $\beta\uparrow\infty$. The detailed discussion on the corresponding thermodynamics system in the replica analysis can be found in \cite{bereyhi2016statistical} and the references therein. For sake of brevity, we skip the details and determine the entropy for the thermodynamic system corresponding to LSE precoders by following the approach in \cite{zaidel2012vector}. In this case, the zero-temperature entropy for our setup is determined as
 \begin{equation}
\mathcal{H}_{0}= \zeta\R_{\mt{R}}(-\zeta)-\int_{0}^\zeta,
\R_{\mt{R}}(-w)\dif w, \label{eq:ZTEnt}
 \end{equation}
where $\zeta=\chi$ under the RS assumption and $\zeta=\chi_1$ under $1$-RSB. When the solution under the assumed structure is exact, the parameter $\mathcal{H}_0$ is zero. Nevertheless, for the cases in which the assumed structure does not lead to the exact solution, $\mathcal{H}_0$ may take some negative values. In these cases, the value of $\mathcal{H}_0$ indicates the accuracy of the solution. The closer $\mathcal{H}_0$ to zero, the better accuracy has the replica solution under the assumed structure.

\section{Special Forms of LSE Precoders}
\label{sec:RS_special}

Using the results given in Section \ref{sec:large}, we derive the asymptotic distortion for some special forms of LSE precoders which address several signal constraints in massive MIMO transmitters. For these forms, we calculate the replica predicted distortion under the RS assumption. The solution under the 1-RSB assumption, however, is only considered in the numerical results section since the derivations in this case are more complicated. 
Throughout our investigations in this section, we consider the channel matrix to be a $K\times N$ iid matrix\footnote{Here, the results do not depend on the distribution of the entries and assuming matrix to be iid is sufficient.} whose entries are zero-mean with variance ${1}/{N}$. This channel model holds in rich scattering environments when perfect power control is employed at user terminals. The results in Proposition~\ref{prop1} and \ref{prop2} are however for a general $\mt H$ and can be used for other channel models as well. In Appendix~\ref{pathlosseff}, we discuss the effect of path-loss in cases with no power control. Considering the iid channel matrix, the $\rm R$-transform of the matrix $\mt R$ in this case reads \cite{tulino2004random}
\begin{eqnarray}
\R_{\mt{R}}(w)=\frac{\alpha^{-1}}{1- w }. \label{eq:r-trans-iid}
\end{eqnarray}
By setting different forms of $\mathbb X$, the asymptotic distortion under the RS and $1$-RSB assumptions can be derived from Proposition~\ref{prop1} and \ref{prop2} using \eqref{eq:r-trans-iid}. In the sequel, we determine the asymptotic distortion under RS  explicitly for some given cases.

\subsection{Per-antenna Peak Power Constraint} 
Using the LSE precoders, the instantaneous power on each transmit antenna at the base station can be constrained by setting $\mathbb X$ to be
\begin{equation}
\mathbbmss{X}=\left\{x=r\e^{j\theta} :\hspace*{.5mm} \theta\in\left[0,2\pi \right] \hspace*{3mm} \text{and} \hspace*{3mm} 0\leq r\leq \sqrt{P}\right\}.
\end{equation}
Here, the peak power of each precoded symbol is constrained to be less than $P$. The tuning factor $\lambda$ in the LSE precoder, moreover, restricts the average transmit power which along with the peak power constraint can limit the transmit PAPR. Using Proposition \ref{prop2}, the RS fixed-point equations in this case reduce to
\begin{subequations}
\begin{eqnarray}\label{peakceq1}
\chi &=& h \left( 1+\chi \right) \sqrt{\frac{\alpha}{q+\gamma\sigma_u^2}},  \\
q&=&c^2 \left[ 1-\e^{-{P}/{c^2}} \right],
\end{eqnarray}
\end{subequations}
where the scalars $c$ and $h$ are defined as
\begin{subequations}
\begin{align}\label{peakceq2}
c &= \dfrac{\sqrt{\alpha \left( q+\gamma\sigma_u^2 \right)}}{\alpha \lambda \left(1+\chi \right)+1}, \\
h
&= c \left[ 1- \e^{-{P}/{c^2}} \right]+\sqrt{P \pi} \hspace*{.3mm} {\Q} (\frac{\sqrt{2P}}{c} ).
\end{align}
\end{subequations}
Moreover, the RS asymptotic distortion for this setup is determined by
\begin{eqnarray} \label{avedistpeak}
D_{\rm RS}=
\frac{q+\gamma\sigma^2_u }{\left( 1+\chi \right)^2}.
\end{eqnarray}
By tuning $\lambda$ properly, the average power, and thus, the transmit PAPR is constrained. The factors $q$ and $\chi$ can then be determined from the fixed-point equations and the distortion under the RS assumption is found from \eqref{avedistpeak}. We later discuss the tuning strategy in Section \ref{sec:optimal}.
%
%

\subsection{$M$-PSK Signals on Antennas}
The LSE precoded signal can be forced to be taken from the $M$-PSK constellation by setting
\begin{eqnarray}\label{pskset}
\mathbbmss{X}=\left\lbrace \sqrt{p}\hspace*{.3mm}\e^{\hspace*{.2mm} j\frac{2\pi}{M}},\sqrt{p}\hspace*{.3mm} \e^{\hspace*{.2mm} j\frac{4\pi}{M}},\sqrt{p}\hspace*{.3mm}\e^{\hspace*{.2mm} j\frac{6\pi}{M}},\cdots, \sqrt{p}\hspace*{.3mm}\e^{\hspace*{.2mm} j2\pi}\right\rbrace.
\end{eqnarray} 
This limitation appears in load-modulated single-RF MIMO transmitters in which due to the limited number of states, the signal constellation is restricted to a finite set of points. Considering the LSE precoder in this case, one observes that $\|\matr x \|^2$ is constant for PSK constellations. Consequently, the penalty term $\lambda \| \matr x \|^2$ in the minimization \eqref{precrule} is ineffective and can be dropped, i.e., $\lambda=0$. By determining the RS fixed-points, the factor $q$ is determined by $q=p$ which equals to the transmit power. The factor $\chi$ is moreover calculated as
\begin{eqnarray}\label{pskchi}
\chi&=& \left[ \frac{2}{M \sin(\pi/M)}\sqrt{\pi \frac{q+\gamma\sigma^2_u}{ q\alpha}}-1 \right]^{-1}
\end{eqnarray}
and the asymptotic distortion under RS is given by \eqref{avedistpeak} when $q=p$ and $\chi$ is given by \eqref{pskchi}.

Using the results for $M$-PSK constellation, the asymptotic distortion of the LSE precoder with constant-envelope signal on each transmit antenna can be easily obtained by taking the limit $M\uparrow \infty$. In this case, $q=p$ and $\chi$ is determined as
\begin{eqnarray}
\chi&=&\left[ 2\sqrt{\frac{q+\gamma\sigma^2_u}{\pi q \alpha}}-1 \right]^{-1} .
\end{eqnarray}

Considering the RS solution for the $M$-PSK constellation, it is observed that there exists a finite $\alpha^*$ in which for the limit $\alpha\rightarrow \alpha^*$, $\chi$ grows to infinity and the asymptotic distortion converges to zero. This observation contradicts our intuition which conjectures that the distortion for a finite antenna-to-user factor cannot reduce to zero. We justify this conjecture in Appendix~\ref{appen3} by deriving a rigorous lower bound on the asymptotic distortion of $M$-PSK signals. This latter result indicates that the derived distortion under the RS assumption is not exact and given a lower bound on the asymptotic distortion which is not necessarily tight. Consequently, throughout the numerical investigations, we calculate further the distortion under the $1$-RSB assumption for $M$-PSK signals. The simulations show that the solution under $1$-RSB assumption gives a tighter lower bound on the asymptotic distortion. Detailed discussions are later presented in Section \ref{sec:num}. 

\section{Tuning Strategy for LSE Precoders} 
\label{sec:optimal}
For each signal constraint, the LSE precoder can be tuned with the parameters $\lambda$ and $\gamma$. The parameter $\lambda$ controls the average transmit power, and $\gamma$ is the power gain of users' signals at the receive side. These parameters affect the transmit power and need to be tuned properly for some given transmit power constraint. In this section, we illustrate the tuning strategy for these parameters and derive their optimal values. To do so, we start by stating the relation between the average transmit power and the replica solutions. 

As it has been observed for the $M$-PSK constellation, the scalar $q$ in the RS solution determines the average transmit power $p$. It is shown that this result holds in general which means that the scalar $q$ in the RS solution determines the average transmit power under the RS assumption \cite{zaidel2012vector}. Under the $1$-RSB assumption, it is $q_1+p_1$ which determined the average power \cite{zaidel2012vector}. This connection between the replica solutions and the transmit average power enables us to tune $\lambda$ and $\gamma$ according to the average power constraint. In fact for the RS solution, by setting the scalar $q$ to be the average power constraint, we have the variables $\chi$, $\gamma$ and $\lambda$ and two fixed-point equations. In this case, one can determine the factor $\lambda$ in terms of $\gamma$, and then, tune $\gamma$ such that some given metric is optimized. An example for the metric is the average ergodic rate. With respect to this metric, the factor $\gamma$ is tuned to maximize the lower bound on the average ergodic rate. Same approach can be taken under the $1$-RSB assumption. 

In the following, we consider the example of RZF precoding as a special case of LSE precoders. For this precoder, we discuss the tuning strategy in details. One should note that RS for the RZF precoder is a valid assumption, due to the fact that the optimization problem in this case is convex. The validity of the assumption can be further checked, as the closed-form asymptotic distortion for the RZF precoder has been derived in the literature using random matrix theory. For other form of LSE precoders, we illustrate the tuning strategy throughout the numerical investigations.

\subsection{Tuning Strategy for the RZF Precoder}
In \cite{peel2005vector}, the problem of tuning the regularization factor of RZF precoding has been considered where the authors have found an approximation of optimum $\lambda$. Using the replica solutions, we can address this problem efficiently and find the optimum choices of $\lambda$ and $\gamma$ analytically. For the RZF precoder, we have $\mathbbmss{X}=\mathbbmss{C}$; therefore, the fixed-point equations read
\begin{subequations}
\begin{eqnarray}
q&=&\frac{\alpha(q+\gamma \sigma_u^2)}{[1+\lambda \alpha(1+\chi)]^2}, \\
\chi &=& \frac{\alpha(1+\chi)}{1+\lambda \alpha(1+\chi)}.
\end{eqnarray}
\end{subequations}
Here, the scalar $q$ represents the average transmit power and is fixed. To tune the precoder, we consider $\chi$ to be the free variable and obtain all the other parameters as functions of $\chi$. Thus, 
\begin{subequations}
\begin{eqnarray}
\label{lambdachi}
\lambda &=&\frac{1}{\chi}-\frac{\alpha^{-1}}{1+\chi}\\
\gamma &=&\frac{q}{\sigma_u^2}\left[\frac{\alpha(1+\chi)^2}{\chi^2}-1  \right]
\end{eqnarray}
\end{subequations}
and $D_{\rm RS} = \alpha q \chi^{-2}$. We now invoke Lemma~\ref{khodlemmalowerrate} to bound the average ergodic rate as
\begin{eqnarray}\label{sumratesimplified}
\tilde{R} &\geq& \log\left(\frac{\alpha q(1+\chi)^2- q \chi^2}{\sigma_n^2 \chi^2+\alpha {q} }\right). 
\end{eqnarray}
We wish to optimize the performance of the precoder with respect to $\tilde R$. Hence, we maximize the right hand side of \eqref{sumratesimplified}. By defining $s\define (\alpha-1) {q}/{\sigma_n^2}-1$, it is straightforward to show that the lower bound has a unique local maximum for $\chi\geq 0$ at 
\begin{eqnarray}\label{chiopt}
\chi_{\rm opt}=\frac{1}{2}\left(s+\sqrt{s^2+4\alpha}  \right).
\end{eqnarray}
Consequently, the optimum value for $\lambda$ is obtained by substituting \eqref{chiopt} in \eqref{lambdachi} which reads
\begin{align}
\lambda_{\rm opt}=\frac{2}{s+\sqrt{s^2+4\alpha}}-\frac{2}{\alpha\left(1+s+\sqrt{s^2+4\alpha} \right)}.
\end{align}
A same procedure can be considered for other choices of $\mathbbmss{X}$ to derive $\lambda_{\rm opt}$. Throughout the numerical investigations, we employ this strategy for the LSE precoders with per-antenna peak power and constant envelope constraints, as well as the $M$-PSK output constellation.

\section{Numerical Investigations}
\label{sec:num}
In this section, we numerically investigate the performance of the LSE precoder considered in the last section. For this aim, two performance measures have been considered: the asymptotic distortion $D$ and the average ergodic rate $\tilde R$. For the latter measure, we use the lower bound derived in Lemma~\ref{khodlemmalowerrate}. Throughout the investigations, we set without loss of generality $\sigma_u^2=1$.

\subsection{Per-antenna Peak Power Constraint}
For the case with per-antenna peak power constraints, the optimization problem in \eqref{precrule} is convex. As we mentioned before, it is generally believed in the literature that for such a case the RS assumption is valid. We take the RS solution for the asymptotic distortion. Noting that the simulation of the precoder in this case is computationally feasible, we check the validity of this assumption by comparing the solution with the simulation results. 

\begin{figure}[t!]
\centering
\resizebox{.8\linewidth}{!}{
\pstool[width=0.6\linewidth]{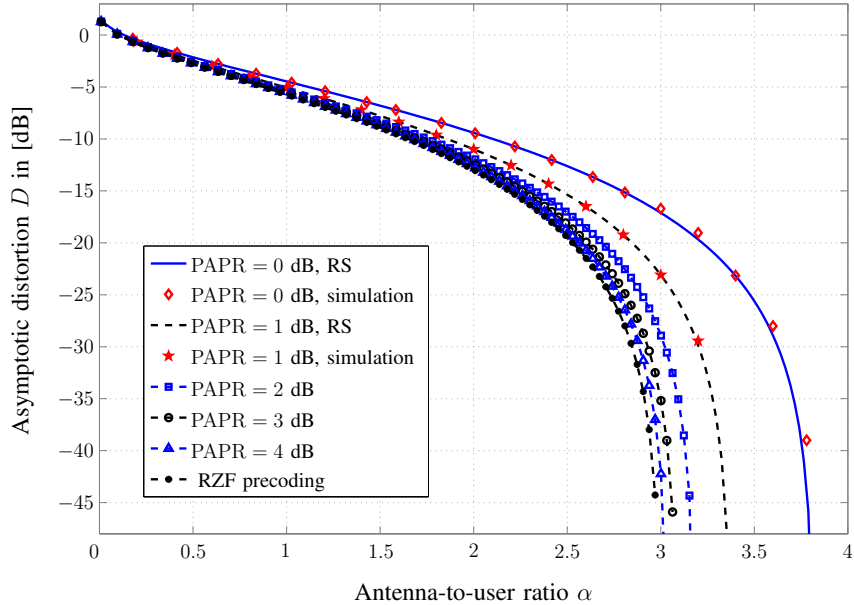}{
\psfrag{average00000000000}[l][l][.5]{$\mathrm{PAPR}=0$ dB, RS}
\psfrag{average11111111000}[l][l][.5]{$\mathrm{PAPR}=0$ dB, simulation}
\psfrag{average22222222000}[l][l][.5]{$\mathrm{PAPR}=1$ dB, RS}
\psfrag{average33333333000}[l][l][.5]{$\mathrm{PAPR}=1$ dB, simulation}
\psfrag{average44444444000}[l][l][.5]{$\mathrm{PAPR}=2$ dB}
\psfrag{average55555555000}[l][l][.5]{$\mathrm{PAPR}=3$ dB}
\psfrag{average66666666000}[l][l][.5]{$\mathrm{PAPR}=4$ dB}
\psfrag{average77777777000}[l][l][.5]{ RZF precoding}
\psfrag{Asymptoticdistortion}[c][b][.6]{Asymptotic distortion $D$ in [dB]}
\psfrag{alpha}[c][t][.6]{Antenna-to-user ratio $\alpha$}

\psfrag{0}[c][l][.5]{$0$}
\psfrag{0.5}[c][c][.5]{$0.5$}
\psfrag{1}[c][c][.5]{$1$}
\psfrag{1.5}[c][c][.5]{$1.5$}
\psfrag{2}[c][c][.5]{$2$}
\psfrag{2.5}[c][c][.5]{$2.5$}
\psfrag{3}[c][c][.5]{$3$}
\psfrag{3.5}[c][c][.5]{$3.5$}
\psfrag{4}[c][c][.5]{$4$}

\psfrag{-5}[c][c][.5]{$-5$}
\psfrag{-10}[c][c][.5]{$-10$}
\psfrag{-15}[c][c][.5]{$-15$}
\psfrag{-20}[c][c][.5]{$-20$}
\psfrag{-25}[c][c][.5]{$-25$}
\psfrag{-30}[c][c][.5]{$-30$}
\psfrag{-35}[c][c][.5]{$-35$}
\psfrag{-40}[c][c][.5]{$-40$}
\psfrag{-45}[c][c][.5]{$-45$}
\psfrag{-50}[c][c][.45]{$-50$}

}}
\caption{Asymptotic distortion versus the antenna-to-user ratio, i.e., $\alpha=N/K$, for several per-antenna peak power constraints when the average transmit power is set to $q=0.5$.}

\label{peak_average_fix_q_difflam}
\end{figure}

For this case, we have two constraints: 
\begin{inparaenum}
\item the total average power should be less than $q$, and
\item the per-antenna peak power should be less than $P$.
\end{inparaenum}
Let $\mathrm{PAPR} \define P/q$ be the PAPR of the precoded signal. Fig.~\ref{peak_average_fix_q_difflam} represents the asymptotic distortion under the RS assumption versus the antenna-to-user ratio $\alpha$ for a fixed total average power $q=0.5$ and $\gamma=1$. To validate the results by the replica method, we have also plotted simulation results obtained by the CVX toolbox for $K=200$. From Fig.~\ref{peak_average_fix_q_difflam}, it is observed that the simulation results confirm the validity of the RS prediction. Moreover, the figure shows that, for the PAPRs more than $3~{\rm dB}$, the asymptotic distortion is sufficiently close to the case without peak power constraint which is in fact the RZF precoding scheme. The curve for ${\rm PAPR}=0~{\rm dB}$ is also plotted which describes the performance of the LSE precoder with constant envelope output signals. The LSE precoder rule in this case is not a convex optimization problem, and thus, it is not guaranteed that the replica method under RS gives the exact distortion. Hence, further studies under RSB are needed to be considered. As the numerical investigations show the consistency of the RS solution with the simulations, we skip further discussions under RSB and leave it as a possible future study.
%
%

\begin{figure}[t]
\centering
\resizebox{.8\linewidth}{!}{
\pstool[width=0.6\linewidth]{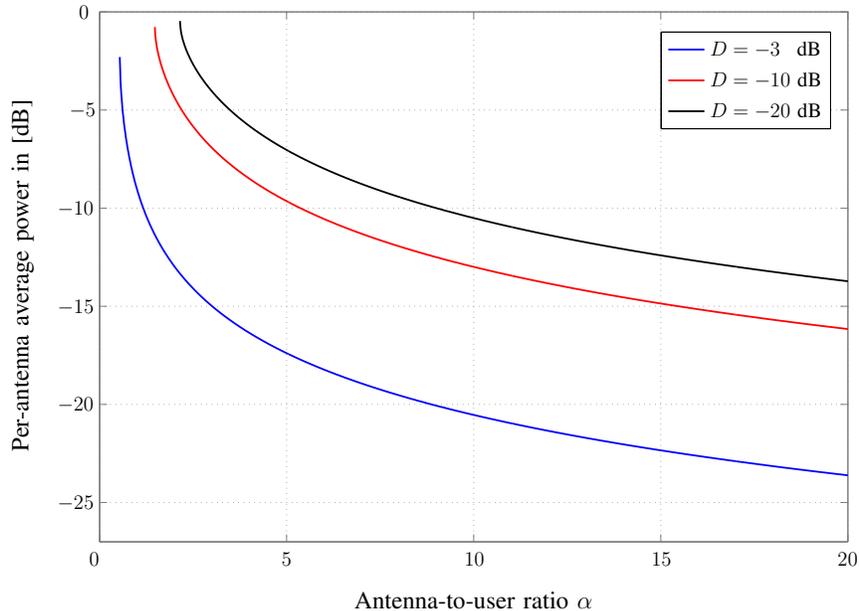}{
\psfrag{Per-antenna average power (dB)}[c][c][.8]{Per-antenna average power [dB]}
\psfrag{D-003DBB}[l][l][.5]{$D=-3$ \hspace{.5mm} dB}
\psfrag{D-010DBB}[l][l][.5]{$D=-10$ dB}
\psfrag{D-020DBB}[l][l][.5]{$D=-20$ dB}
\psfrag{Per-antenna average power (dB)}[c][b][.6]{Per-antenna average power in [dB]}
\psfrag{alpha}[c][t][.6]{Antenna-to-user ratio $\alpha$}

\psfrag{0}[c][l][.5]{$0$}
\psfrag{5}[c][c][.5]{$5$}
\psfrag{10}[c][c][.5]{$10$}
\psfrag{15}[c][c][.5]{$15$}
\psfrag{20}[c][c][.5]{$20$}
\psfrag{2.5}[c][c][.5]{$2.5$}
\psfrag{3}[c][c][.5]{$3$}
\psfrag{3.5}[c][c][.5]{$3.5$}
\psfrag{4}[c][c][.5]{$4$}

\psfrag{-5}[c][c][.5]{$-5$}
\psfrag{-10}[c][c][.5]{$-10$}
\psfrag{-15}[c][c][.5]{$-15$}
\psfrag{-20}[c][c][.5]{$-20$}
\psfrag{-25}[c][c][.5]{$-25$}
\psfrag{-30}[c][c][.5]{$-30$}
\psfrag{-35}[c][c][.5]{$-35$}
\psfrag{-40}[c][c][.5]{$-40$}
\psfrag{-45}[c][c][.5]{$-45$}
\psfrag{-50}[c][c][.45]{$-50$}

}}
\caption{Required per-antenna average power versus the antenna-to-user ratio, i.e., $\alpha=N/K$, for various asymptotic distortions when the peak power is set to $P=1$.}
\label{average_vs_alpha_fixedD}
\end{figure}

We now study the variation of the required average transmit power for a given asymptotic distortion with respect to the number of transmit antennas. For this goal, we consider the case with unit per-antenna peak power constraint, i.e., $P=1$, and plot the average power per-antenna in terms of $\alpha$ for several given asymptotic distortions. The results are shown in Fig.~\ref{average_vs_alpha_fixedD}. Here, the parameter $\gamma$ is set to $1$. As the figure depicts, the per-antenna average power decays by increasing $\alpha$. By numerical curve fitting, it can be observed that the per-antenna average power asymptotically\footnote{Here, by asymptotically we mean as $\alpha$ grows large.} decays in the form of $c\hspace*{.3mm}\alpha ^{\kappa}$ for some constant $c$ and $\kappa=-1$. For non-asymptotic regime of $\alpha$, however, $\kappa<-1$. This observation agrees with the earlier results reported in the literature which indicates that for massive MIMO systems with average power constraint the signal to interference and noise ratio can be improved by a factor of $\alpha$ asymptotically when the channel state information is perfectly available at the base station \cite{rusek2013scaling}.

\begin{figure}[t]
\centering
\resizebox{.8\linewidth}{!}{
\pstool[width=0.6\linewidth]{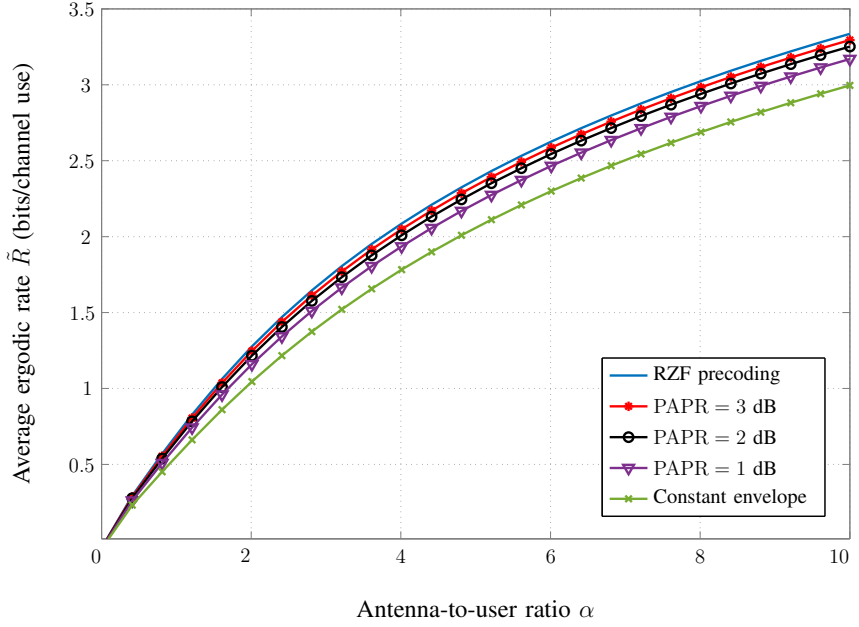}{
\psfrag{Achievable rate (bits/channel use)}[c][c][.6]{Average ergodic rate $\tilde{R}$ (bits/channel use)}
\psfrag{alpha}[c][c][.6]{Antenna-to-user ratio $\alpha$}

\psfrag{DDDDDDBAA}[l][l][.5]{RZF precoding}
\psfrag{DDDDDDBBB}[l][l][.5]{$\mathrm{PAPR}=3$ dB}
\psfrag{DDDDDDBCC}[l][l][.5]{$\mathrm{PAPR}=2$ dB}
\psfrag{DDDDDDBDD}[l][l][.5]{$\mathrm{PAPR}=1$ dB}
\psfrag{DDDDDDBEE}[l][l][.5]{Constant envelope}

\psfrag{0}[c][l][.5]{$0$}
\psfrag{2}[c][c][.5]{$2$}
\psfrag{4}[c][c][.5]{$4$}
\psfrag{6}[c][c][.5]{$6$}
\psfrag{8}[c][c][.5]{$8$}
\psfrag{10}[c][l][.5]{$10$}
\psfrag{1}[c][l][.5]{$1$}
\psfrag{0.5}[c][c][.5]{$0.5$}
\psfrag{1.5}[c][c][.5]{$1.5$}
\psfrag{2}[c][l][.5]{$2$}
\psfrag{2.5}[c][c][.5]{$2.5$}
\psfrag{3}[c][l][.5]{$3$}
\psfrag{3.5}[c][c][.5]{$3.5$}
}}
\caption{The average ergodic rate versus the antenna-to-user ratio $\alpha$ for several PAPR constraints when the average transmit power is set to $q=1$ and $\sigma_n^2=1$. $\gamma$ and $\lambda$ are tuned to fulfill the power constraint and to maximize~the~rate.}
\label{rate}
\end{figure}

We further investigate the lower bound on the average ergodic rate using Lemma~\ref{khodlemmalowerrate}. We set the noise variance $\sigma_n^2$ and the average transmitted power $q$ to $1$ and choose the parameters $\lambda$ and $\gamma$ such that the lower bound is maximized. Fig.~\ref{rate} shows the lower bound as a function of the antenna-to-user ratio $\alpha$ for several PAPRs. One should note that although the results given by the replica method also predicts the distortion for $\alpha<1$, the desired region is $\alpha\geq1$. This is due to the fact that the number of base station antennas are usually larger than the number of users. As it is observed in the figure, the average ergodic rate for several PAPRs are close. Here, the LSE precoder with constant envelope signals show tha most degraded performance. For this precoder at $\alpha=5$, we need about $20\%$ more antennas to achieve the performance given by the LSE precoder which is not constrained in terms of PAPR, i.e., RZF precoder.

\begin{figure}[t]
\centering
\resizebox{.8\linewidth}{!}{
\pstool[width=0.6\linewidth]{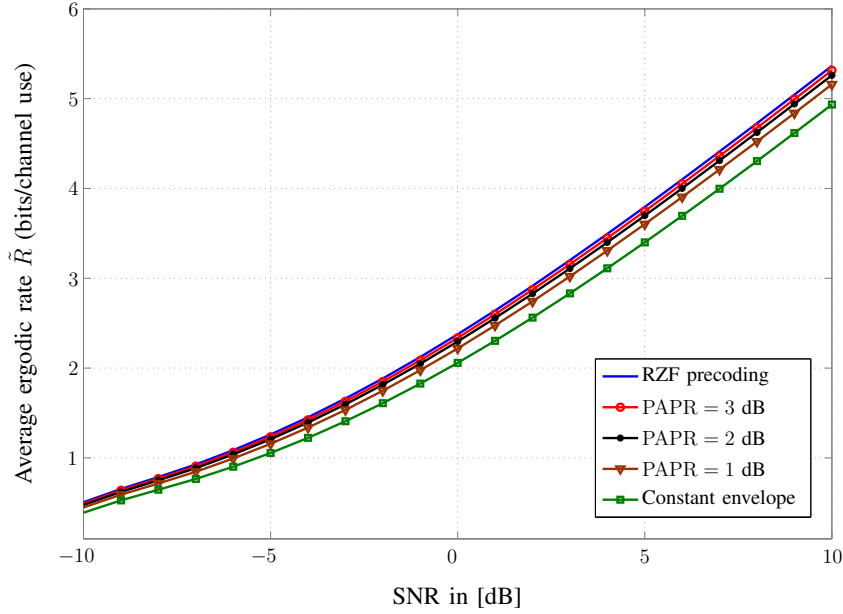}{
\psfrag{Achievable rate (bits/channel use)}[c][c][.6]{Average ergodic rate $\tilde{R}$ (bits/channel use)}
\psfrag{SNR (dB)}[c][t][.6]{SNR in [dB]}

\psfrag{AAABBBKKAA}[l][l][.5]{RZF precoding}
\psfrag{AAABBBKKBB}[l][l][.5]{$\mathrm{PAPR}=3$ dB}
\psfrag{AAABBBKKCC}[l][l][.5]{$\mathrm{PAPR}=2$ dB}
\psfrag{AAABBBKKDD}[l][l][.5]{$\mathrm{PAPR}=1$ dB}
\psfrag{AAABBBKKEE}[l][l][.5]{Constant envelope}

\psfrag{0}[c][l][.5]{$0$}
\psfrag{1}[c][c][.5]{$1$}
\psfrag{1.5}[c][c][.5]{$1.5$}
\psfrag{2}[c][c][.5]{$2$}
\psfrag{2.5}[c][c][.5]{$2.5$}
\psfrag{3}[c][c][.5]{$3$}
\psfrag{3.5}[c][c][.5]{$3.5$}
\psfrag{4}[c][c][.5]{$4$}
\psfrag{4.5}[c][c][.5]{$4.5$}
\psfrag{5}[c][c][.5]{$5$}
\psfrag{5.5}[c][c][.5]{$5.5$}
\psfrag{-10}[c][c][.5]{$-10$}
\psfrag{6}[c][c][.5]{$6$}
\psfrag{-5}[c][c][.5]{$-5$}
\psfrag{10}[c][c][.5]{$10$}
}}
\caption{The average ergodic rate versus the SNR for different PAPR when $q=1$ and $\alpha=5$. The parameters $\gamma$ and $\lambda$ are tuned such that the power constraint is fulfilled and the rate is maximized.}
\label{rate_SNR}
\end{figure}

To study the achievable average ergodic rate at different noise powers, we plot the lower bound on $\tilde R$ at $\alpha=5$ 
in terms of the Signal-to-Noise Ratio (SNR) in Fig.~\ref{rate_SNR}. Here, the SNR is defined to be ${q}/{\sigma_n^2}$ and the average power constraint is set to $q=1$. Other parameters are calculated such that the lower bound is optimized. The figure shows that at $\alpha=5$ about $1.3~{\rm dB}$ more transmit power is required to get the same performance compared to the RZF precoder. 

\subsection{$M$-PSK signals}
\begin{figure}[t!]
\centering
\resizebox{.8\linewidth}{!}{
\pstool[width=.6\linewidth]{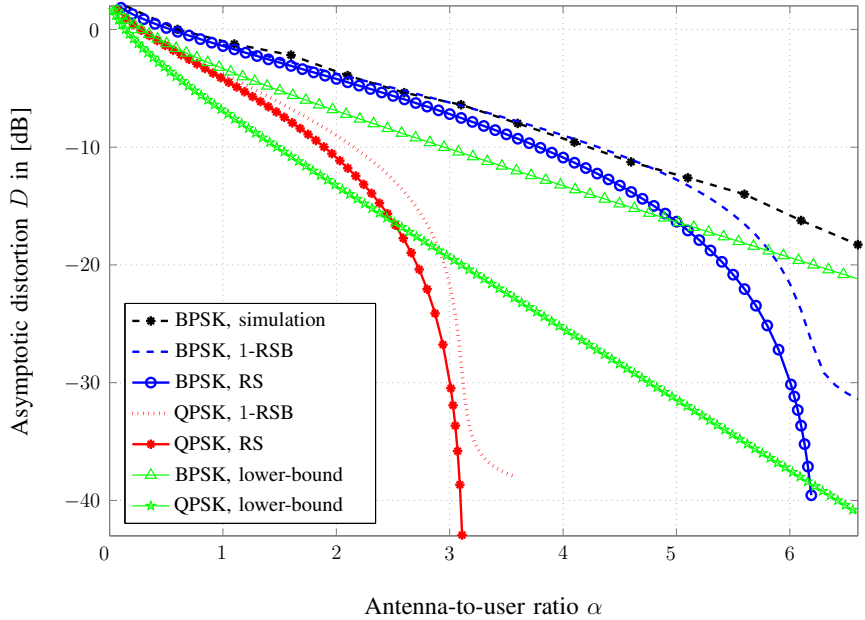}{
\psfrag{Average distortion (dB)}[c][c][.6]{Asymptotic distortion $D$ in [dB]}
\psfrag{alpha}[c][c][.6]{Antenna-to-user ratio $\alpha$}

\psfrag{BPSK-RSBSAAA}[l][l][.5]{BPSK, simulation}
\psfrag{BPSK-RSBBAAA}[l][l][.5]{BPSK, $1$-RSB}
\psfrag{BPSK-RSBNAAA}[l][l][.5]{BPSK, RS}
\psfrag{QPSK-RSBBAAA}[l][l][.5]{QPSK, $1$-RSB}
\psfrag{QPSK-RSBNAAA}[l][l][.5]{QPSK, RS}
\psfrag{BPSK-RSBLAAA}[l][l][.5]{BPSK, lower-bound}
\psfrag{QPSK-RSBLAAA}[l][l][.5]{QPSK, lower-bound}

\psfrag{0}[c][l][.5]{$0$}
\psfrag{5}[c][l][.5]{$5$}
\psfrag{10}[c][l][.5]{$10$}
\psfrag{15}[c][l][.5]{$15$}
\psfrag{20}[c][l][.5]{$20$}
\psfrag{-25}[c][l][.5]{$-25$}
\psfrag{-5}[c][c][.5]{$-5~$}
\psfrag{-10}[c][c][.5]{$-10~$}
\psfrag{-15}[c][c][.5]{$-15~$}
\psfrag{-20}[c][c][.5]{$-20~$}
\psfrag{-25}[c][c][.5]{$-25~$}
\psfrag{-30}[c][c][.5]{$-30~$}
\psfrag{-35}[c][c][.5]{$-35~$}
\psfrag{-40}[c][c][.5]{$-40~$}
\psfrag{-45}[c][c][.5]{$-45~$}
\psfrag{-50}[c][c][.5]{$-50~$}

\psfrag{1}[c][c][.5]{$1$}
\psfrag{2}[c][c][.5]{$2$}
\psfrag{3}[c][c][.5]{$3$}
\psfrag{4}[c][c][.5]{$4$}
\psfrag{9}[c][c][.5]{$5$}
\psfrag{6}[c][c][.5]{$6$}
\psfrag{7}[c][c][.5]{$7$}
\psfrag{8}[c][c][.5]{$8$}
}}
\caption{Asymptotic distortion as a function of the antenna-to-user ratio, i.e., $\alpha=N/K$, for the LSE precoders with output BPSK and QPSK constellations.}
\label{psksigma1}
\end{figure}

Since the LSE precoder with $M$-PSK constellation at the output deals with a combinatorial problem, one needs to investigate the validity of the RS assumption in this case. As we mentioned in the earlier sections, the exact distortion in this case is not given under the RS assumption, and thus, further investigations under RSB are required. 
Fig.~\ref{psksigma1} shows the asymptotic distortion for the BPSK and QPSK constellations when $q=1$ and $\gamma=1$. For the sake of comparison, a lower bound based on the union bound on the asymptotic distortion is also plotted; see Appendix~\ref{appen3} for the derivation of the lower bound. The numerical simulations, which have been obtained by integer programming, are also shown for the BPSK constellation considering $N=100$. As it is observed for the BPSK constellation, the RS solution starts to deviate from the simulations as $\alpha$ grows. For $\alpha\geq 5$, the asymptotic distortion determined under the RS assumption even violates the lower bound given by the union bound. This observation clarifies that the RS assumption in this case fails to give a tight lower bound on the asymptotic distortion. 
The asymptotic distortion calculated under $1$-RSB gives a tighter lower bound which outperform the union bound within a larger range of antenna-to-user ratios compared to the RS solution. However, similar to the RS solution, the 1-RSB solution starts to deviate the simulations as $\alpha$ grows and deviates the lower bound based on the union bound somewhere near to $\alpha=6$. Tighter lower bounds on asymptotic distortion in this case can be obtained using RSB with higher steps. Considering the figure, similar results are observed for the QPSK constellation.

\begin{figure}[t!]
\centering
\resizebox{.8\linewidth}{!}{
\pstool[width=.6\linewidth]{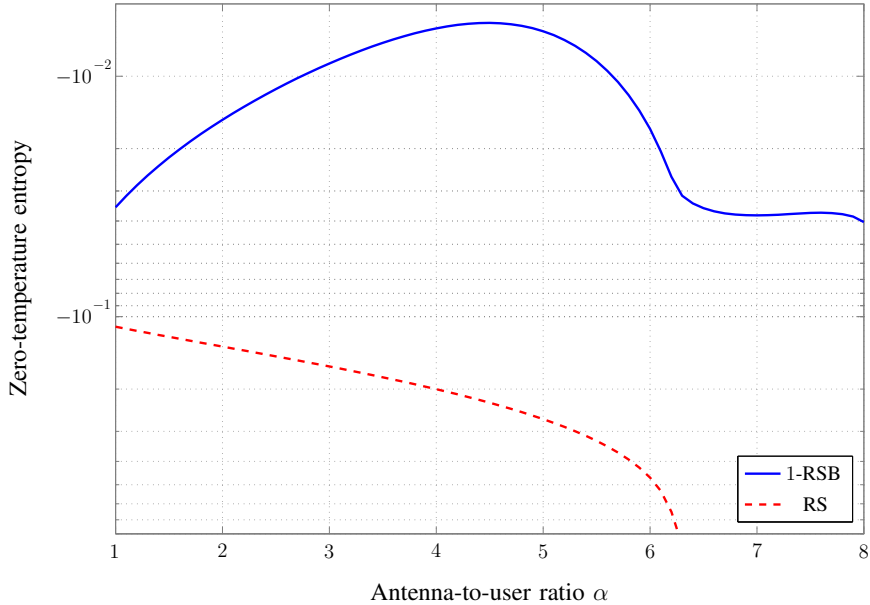}{
\psfrag{Entropy}[c][b][.6]{Zero-temperature entropy}
\psfrag{alpha}[c][t][.6]{Antenna-to-user ratio $\alpha$}
\psfrag{RSB1}[c][c][.5]{$1$-RSB}
\psfrag{RSB0}[c][c][.5]{RS}

\psfrag{8}[c][c][.5]{$8$}
\psfrag{1}[c][c][.5]{$1$}
\psfrag{2}[c][c][.5]{$2$}
\psfrag{3}[c][c][.5]{$3$}
\psfrag{4}[c][c][.5]{$4$}
\psfrag{5}[c][c][.5]{$5$}
\psfrag{6}[c][c][.5]{$6$}
\psfrag{7}[c][c][.5]{$7$}

\psfrag{-10}[c][c][.5]{$-10$}
\psfrag{-1}[c][c][.35]{$-1$}
\psfrag{-2}[c][c][.35]{\hspace*{-.4mm}$-2$}
\psfrag{-3}[c][l][.5]{$-3$}

}}
\caption{Zero-temperature entropy versus the antenna-to-user ratio for BPSK signals.}
\label{Entropy}
\end{figure}

In order to see the reliability of the solutions we invoke the consistency test based on the zero-temperature entropy. Fig.~\ref{Entropy} shows the zero-temperature entropy given in \eqref{eq:ZTEnt} as a function of the antenna-to-user ratio for the BPSK constellation. As it is observed, under the $1$-RSB assumption the zero-temperature entropy is much closer to zero. This observation implies that the $1$-RSB solution as an approximation of the asymptotic distortion is more reliable compared to the RS solution. Moreover, as the zero-temperature entropy in the both cases takes negative values, we can conclude that neither the RS nor the $1$-RSB solution is exact. 

\begin{figure}[t!]
\centering
\resizebox{.8\linewidth}{!}{
\pstool[width=.6\linewidth]{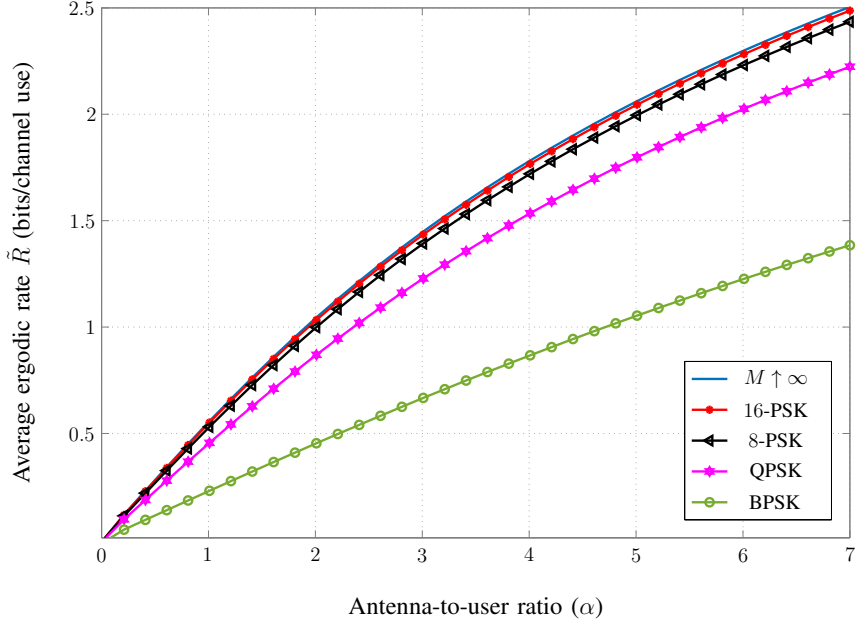}{
\psfrag{alpha}[c][c][.6]{Antenna-to-user ratio ($\alpha$)}
\psfrag{Achievable rate per user (bits/channel use)}[c][c][.6]{Average ergodic rate $\tilde{R}$ (bits/channel use)}
\psfrag{GGMML}[c][c][.5]{$M\uparrow \infty$}
\psfrag{A6PSKL}[c][c][.5]{$16$-PSK}
\psfrag{A8PSKL}[c][c][.5]{$8$-PSK}
\psfrag{AQPSK1}[c][c][.5]{QPSK}
\psfrag{ABPSK1}[c][c][.5]{BPSK}

\psfrag{0}[c][c][.5]{$0$}
\psfrag{1}[c][c][.5]{$1$}
\psfrag{2}[c][c][.5]{$2$}
\psfrag{3}[c][c][.5]{$3$}
\psfrag{4}[c][c][.5]{$4$}
\psfrag{5}[c][c][.5]{$5$}
\psfrag{6}[c][c][.5]{$6$}
\psfrag{7}[c][c][.5]{$7$}

\psfrag{0.5}[c][c][.5]{$0.5$}
\psfrag{1.5}[c][c][.5]{$1.5$}
\psfrag{2.5}[c][c][.5]{$2.5$}

}}
\caption{The lower-bound on the average ergodic rate in terms of the antenna-to-user ratio under the RS assumption for several PSK schemes. $\gamma$ is tuned such that the bound is maximized.}
\label{RS_psk_rate}
\end{figure}

Fig.~\ref{RS_psk_rate} shows the variations of the average ergodic rate versus the antenna-to-user ratio $\alpha$ for the LSE precoder with $M$-PSK constellations under the RS assumption. Here, the lower bound in Lemma~\ref{khodlemmalowerrate} is considered, and $\gamma$ is tuned to maximize the lower bound. For $M\geq 3$ the curves start to lie very close to average ergodic rate of the constant envelope precoder. By reminding the fact that for the constant envelope case the RS solution is exact, one can conclude that in MIMO transmitters with LSE precoders an acceptable performance can be obtained by use of $8$-PSK constellation instead of the whole complex unit circle.  
 
\section{Conclusions}
\label{sec:conclusion}
In this paper, we studied the large-system performance of a class of nonlinear Least Square Error (LSE) precoders in massive MIMO systems. This class of precoders designs the transmit signals such that the distortion at the user terminals is minimized subject to some given constraints on the output signals. By invoking the replica method from statistical physics, we derived the asymptotic distortion under the Replica Symmetry (RS) and the Replica Symmetry Breaking (RSB) assumptions. The analytical results were then investigated for some special forms of LSE precoders; namely, the precoders with limited output peak power, constant envelope signals and finite output constellation.Our numerical investigations showed that in the case with peak power constraint, the analytical formula derived by the replica method is perfectly consistent with the simulations. For $M$-PSK constellations, however, the replica analysis give a lower-bound on the asymptotic distortion. The lower-bound in this case was shown to be tighter under the $1$-RSB assumption compared to that given by RS. Nevertheless, the $1$-RSB solution is also not exact in this case. It is therefore required to consider the RSB solutions with more steps to assess the exact performance of the LSE precoders with finite output constellations in the large-system limit. Using the large-system results, we further proposed a tuning strategy for the LSE precoders. The tuning strategy enabled us to derive the optimum regularization factor for the Regularized Zero Forcing (RZF) precoder in closed-form.

Regarding the LSE precoders, there are still several questions to be answered. For example, one may study the joint distribution of a given user's data and its corresponding signal received at the user terminal after the precoding and transmission. The joint distribution could then lead us to derive the exact average ergodic rate achieved in a broadcast MIMO channel when an LSE precoder is employed. Another interesting direction is to study the asymptotic decoupling principle, shown in \cite{guo2005randomly} for multiuser CDMA systems, for this setup. Using the replica solutions given in this paper, one may further study the optimal choice for the constellation set $\mathbbmss{X}$ of LSE precoding when the number of signal points are constrained to be fix. Finally, practical and feasible algorithms for implementation of LSE precoders need to be investigated.

\begin{appendices}
\newpage
\section{Proof of Lemma~\ref{khodlemmalowerrate}}\label{lemmalowerrate}
Using the definition of the mutual information, the average ergodic rate can be written as 
\begin{eqnarray}\label{lowerbdovom1a}
\tilde{R}&=&\frac{1}{K}\expect\limits_{\mt{H}}\sum_{i=k}^K {\rm I}(y_k;u_k)=
\frac{1}{K}\expect\limits_{\mt{H}}\sum_{k=1}^K {\rm h}(u_k)-{\rm h}(u_k|y_k).
\end{eqnarray}
One can use the equality 
\begin{eqnarray}
{\rm h}(X|Y)={\rm h}(X-cY|Y)
\end{eqnarray}
for any constant $c$ and write \eqref{lowerbdovom1a} as
\begin{eqnarray}\tilde{R}=
\frac{1}{K}\expect\limits_{\mt{H}}\sum_{k=1}^K {\rm h}(u_k)-{\rm h}\left(u_k-\frac{1}{\sqrt{\gamma}}y_k\big| y_k\right)=
\frac{1}{K}\expect\limits_{\mt{H}}\sum_{k=1}^K {\rm h}(u_k)-{\rm h}\left(\frac{z_k(\mt{H},\matr{u})+n_k}{\sqrt{\gamma}} \big| y_k \right).
\end{eqnarray}
Next using the inequality
\begin{eqnarray}
{\rm h}(X)\geq{\rm h}(X|Y),
\end{eqnarray}
a lower bound for the average ergodic rate is derived as follows
\begin{eqnarray}
\tilde{R}
&\geq&\frac{1}{K}\expect\limits_{\mt{H}}\sum_{k=1}^K {\rm h}(u_k)-{\rm h}\left(\frac{z_k(\mt{H},\matr{u})+n_k}{\sqrt{\gamma}}\right).
\end{eqnarray}
Since $n_k$ and $z_{k}(\mt{H},\matr{u})$ are independent and the maximum entropy for the interference term is resulted when $z_{k}(\mt{H},\matr{u})$ is Gaussian distributed, we have
\begin{eqnarray}\label{lowerbdovom}
\tilde{R} \geq \frac{1}{K}\sum_{k=1}^K\expect\limits_{\mt{H}}\log\left(\frac{\gamma\sigma_u^2}{\sigma_n^2+\expect\limits_{\matr{u}}|z_k(\mt{H},\matr{u})|^2}\right).
\end{eqnarray}
Using Jensen's inequality and the fact that the function $\log \left( {\gamma\sigma_u^2}/\left[{\sigma_n^2+x}\right] \right)$ is convex, we obtain
\begin{eqnarray}
\tilde{R}\geq \log\left(\frac{\gamma\sigma_u^2}{\sigma_n^2+\frac{1}{K}\sum\limits_{k=1}^K \expect\limits_{\mt{H},\matr{u}} |z_k(\mt{H},\matr{u})|^2}\right).
\end{eqnarray}
From \eqref{orgopt}, one can write $ D=\lim\limits_{K\rightarrow \infty}\frac{1}{K}\sum_{k=1}^K \expect\limits_{\mt{H},\matr{u}}|z_k(\mt{H}, \matr{u})|^2$, and hence
\begin{eqnarray}\label{achievableaver2}
\tilde{R} \geq \log\left(\frac{\gamma\sigma_u^2}{\sigma_n^2+D}\right),
\end{eqnarray}
for ${K\rightarrow \infty}$. \eqref{achievableaver2} concludes the proof.

\section{Proof of Proposition~\ref{prop1}}\label{app1}
We intend to find the matrices $\mt Q$ and $\tilde{\mt Q}$ which satisfy the RS constraint and minimize
\begin{eqnarray} \label{exponent}
\mathcal{G}(\mt{Q})+{\tr}(\tilde{\mt{Q}} \mt{Q}) - \log \mathcal{M} (\tilde{\mt{Q}}).
\end{eqnarray}
Using the RS structure for the matrices $\mt{Q}$ and $\tilde{\mt{Q}}$, we derive
\begin{eqnarray}
\mathcal{G}(\mt{Q})&=&n \lambda(\beta q+\chi)+ (n-1)\int_0^\chi R(-w)\dif w + \int_0^{\gamma_0} R(-w)\dif w,
\end{eqnarray}
where $\gamma_0=\chi - n \chi\beta \gamma\sigma_u^2+nq\beta - n^2 q\beta^2 \gamma\sigma_u^2$. The trace term and $\log \mathcal{M} (\tilde{\mt{Q}})$ are also derived taking the same steps as in \cite{zaidel2012vector}. We therefore skip the detailed derivations of these terms. 
By letting the derivatives with respect to $q$ and $\chi$ to zero we conclude
\begin{subequations}
\begin{eqnarray}
&&n\lambda+\left(n-n^2\beta\gamma \sigma^2_u \right)\R_{\mt{R}}(-\gamma_0)+n(n-1)\beta f^2+n(\beta f^2-e)=0 \label{sadRs1} \\
&&n\lambda+(n-1)\R_{\mt{R}}(-\chi)+\left(1-n\beta \gamma\sigma^2_u\right)\R_{\mt{R}}(-\gamma_0)+n(\beta f^2-e)=0. \label{sadRs2}
\end{eqnarray}
\end{subequations}
Solving \eqref{sadRs1}-\eqref{sadRs2} for $f$ and $e$ and taking the limit $n\downarrow 0$, $f$ and $e$ are determined as
\begin{subequations}
\begin{eqnarray}
f&=&\sqrt{(q-\chi \gamma\sigma_u^2)\R^\prime_\mt{R}(-\chi)+\gamma\sigma^2_u \R(-\chi)} \\
e&=& \R_{\mt{R}}(-\chi)+\lambda.
\end{eqnarray}
\end{subequations}
We further let the derivative of \eqref{exponent} with respect to $f$ and $e$ to be zero which leads us to the following fixed-point equations
\begin{subequations}
\begin{eqnarray}
\chi &=&\frac{1}{f} \int_{\mathbb{C}} \frac{\sum\limits_{x\in \mathbbmss{X}} \Re\{z^*x\}\e^{2\beta f \Re\{ z^*x\}-\beta e |x|^2}  }{\sum\limits_{x\in \mathbbmss{X}} \e^{2\beta f \Re\{ z^*x\}-\beta e |x|^2}} \hspace*{.3mm} \mathrm{D}z \\
q&=&\int_{\mathbb{C}} \frac{\sum\limits_{x\in \mathbbmss{X}} |x|^2\e^{2\beta f \Re\{ z^*x\}-\beta e |x|^2}  }{\sum\limits_{x\in \mathbbmss{X}} \e^{2\beta f \Re\{ z^*x\}-\beta e |x|^2}} \hspace*{.3mm} \mathrm{D}z -\frac{\chi}{\beta}
\end{eqnarray}
\end{subequations}
which by taking the limit $\beta \uparrow \infty $ reduce to
\begin{subequations}
\begin{eqnarray}
\chi&=&\frac{1}{f} \Re \int_{\mathbbmss{C}} \argmin\limits_{x\in \mathbbmss{X}} \left|z-\frac{\R_{\mt{R}}(-\chi)+\lambda }{f}x  \right|z^* \mathrm{D} z \\
q&=&\int_{\mathbbmss{C}} \left|\argmin\limits_{x\in \mathbbmss{X}} \left|z-\frac{\R_{\mt{R}}(-\chi)+\lambda }{f} x  \right| \right|^2 \mathrm{D} z.
\end{eqnarray}
\end{subequations}
Using Lemma~\ref{varahlemma}, the integral in \eqref{replIQ} is replaced with the integrand at the saddle-point, and thus, $\tilde{D}$ under the RS assumption is obtained as
\begin{subequations}
\begin{eqnarray}
\tilde{D}_{\rm RS} &=&\gamma\sigma_u^2+\lim_{\beta\uparrow \infty}\frac{1}{\beta}\lim_{n\downarrow 0}\frac{\partial }{\partial n}\left[  \lambda \alpha n(q\beta +\chi)+\alpha  (n-1)\int_0^{\chi} \R_{\mt{R}}(-w)\dif w +\alpha\int_0^{\gamma_0} \R_{\mt{R}}(-w)\dif w  \right. \nonumber \\
& \ &-\left.\alpha \log(M(e,f))+\alpha n(n-1)f^2\beta^2 q+\alpha n(f^2\beta-e)(\chi+\beta q)\right]  \\
&=&\gamma\sigma_u^2+\lambda \alpha q+\alpha \left[q(\R_{\mt{R}}(-\chi)-\chi \R^{\prime}_{\mt{R}}(-\chi))+\sigma^2_u\chi^2  \R^{\prime}_{\mt{R}}(-\chi)-2\chi \sigma^2_u \R_{\mt{R}}(-\chi)\right]  \\
&=&\gamma\sigma_u^2+\lambda \alpha q+\alpha \frac{\partial}{\partial \chi}\left[ (q-\gamma\sigma^2_u\chi)\chi \R_{\mt{R}}(-\chi)\right],
\end{eqnarray}
\end{subequations}
where we have used the subscript $\rm RS$ to indicate that $\tilde D$ is derived under RS. Consequently, the asymptotic distortion under the RS assumption is determined as $D_{\rm RS}=\tilde{D}_{\rm RS} -\lambda \alpha q$ which reads
\begin{eqnarray}
D_{\rm RS} = \gamma\sigma_u^2+\alpha \frac{\partial}{\partial \chi}\left[ (q-\gamma\sigma^2_u\chi)\chi \R_{\mt{R}}(-\chi)\right]. \label{conc_pro1}
\end{eqnarray}
\eqref{conc_pro1} concludes the proof of Proposition~\ref{prop1}.

\section{Proof of Proposition~\ref{prop2}}\label{rsbproof}
Similar to the RS case, we intend to derive the matrices $\mt Q$ and $\tilde{\mt Q}$ of the form \eqref{RSBstr1} and \eqref{RSBstr2} which maximize the exponent function \eqref{exponent}. We therefore start by determining the exponent function under the $1$-RSB structure. To calculate $\mathcal{G}(\mt{Q})$, one needs to derive the eigenvalues of 
\begin{eqnarray}
\mt{G}
&=& \left(-n\beta\gamma\sigma_u^2 q_1- \gamma\sigma_u^2p_1\mu_1-\gamma\sigma_u^2\chi_1+q_1  \right)\matr{1}_{n\times n}+\frac{\chi_1}{\beta}\mt{I}_{n}+p_1\mt{I}_{\frac{n\beta}{\mu_1}} \otimes \matr{1}_{\tfrac{\mu_1}{\beta}\times \tfrac{\mu_1}{\beta}}
\end{eqnarray}
which is explicitly derived by trivial lines of derivations as in \cite{zaidel2012vector}. Consequently, one obtains
\begin{eqnarray}
\mathcal{G}(\mt{Q})&=&n\lambda (\chi_1 +\beta (q_1+p_1))+\left( n-\frac{n\beta}{\mu_1}\right) \int_0^{\chi_1} \R_{\mt{R}}(-w)\dif w \nonumber \\
& \ & + \left( \frac{n\beta}{\mu_1}-1\right) \int_0^{\eta_1} \R_{\mt{R}}(-w)\dif w + \int_0^{\gamma_1} \R_{\mt{R}}(-w)\dif w .
\end{eqnarray}
where $\eta_1=\chi_1+\mu_1p_1$ and $\gamma_1=\chi_1+\mu_1p_1+\beta n\left(-n\beta\gamma\sigma_u^2 q_1- \gamma\sigma_u^2p_1\mu_1-\gamma\sigma_u^2\chi_1+q_1  \right)$. Furthermore, for the 1-RSB structures  one can determine the trace term as
\begin{eqnarray}
\tr(\tilde{\mt{Q}} \mt{Q})&=&n^2\beta^2f_1^2q_1+n\beta f_1^2p_1 \mu_1+n\beta f_1^2\chi_1+nq_1\beta g_1^2\mu_1+n\beta g_1^2 p_1\mu_1\nonumber \\
& \ & + \hspace*{.7mm} n\beta g_1^2\chi_1-n\beta e_1q_1-n\beta e_1p_1 -n e_1 \chi_1.
\end{eqnarray}
By letting the derivative of the exponent function with respect to $q_1$, $p_1$ and $\chi_1$ be zero,~we~obtain 
\begin{subequations}
\begin{align}
&\lambda \hspace*{-.7mm} + \hspace*{-.7mm} n\beta f_1^2 \hspace*{-.7mm} + \hspace*{-.7mm} \mu_1g_1^2 \hspace*{-.7mm} - \hspace*{-.7mm} e_1  \hspace*{-.7mm} + \hspace*{-.7mm} (1-n\beta\gamma \sigma_u^2)\R_{\mt{R}}(-\gamma_1)=0 \\
&\lambda \hspace*{-.7mm} + \hspace*{-.7mm} \mu_1(f_1^2+g_1^2) \hspace*{-.7mm} - \hspace*{-.7mm} e_1 \hspace*{-.7mm} + \hspace*{-.7mm} \left( 1-\frac{\mu_1}{n\beta}  \right)\R_{\mt{R}}(-\chi_1-\mu_1p_1)+\left(\frac{\mu_1}{n\beta}-\gamma\sigma_u^2 \mu_1  \right)\R_{\mt{R}}(-\gamma_1)=0  \\
&\lambda \hspace*{-.7mm} + \hspace*{-.7mm} \beta f_1^2 \hspace*{-.7mm} + \hspace*{-.7mm} \beta g_1^2 \hspace*{-.7mm} - \hspace*{-.7mm} e_1 \hspace*{-.7mm} +\hspace*{-.7mm} \left( 1 \hspace*{-.7mm} - \hspace*{-.7mm} \frac{\beta}{\mu_1}  \right)\R_{\mt{R}}(-\chi_1) \hspace*{-.8mm} + \hspace*{-.8mm} \left( \frac{\beta}{\mu_1} \hspace*{-.8mm} - \hspace*{-.8mm} \frac{1}{n}\right)\R_{\mt{R}}(-\eta_1) \hspace*{-.7mm} + \hspace*{-.7mm} \frac{1 \hspace*{-.8mm} - \hspace*{-.8mm} \beta n \gamma\sigma_u^2}{n}\R_{\mt{R}}(-\gamma_1) \hspace*{-.8mm} = \hspace*{-.8mm} 0
\end{align}
\end{subequations}
as $K\uparrow \infty$. By solving the equations for $f_1$, $g_1$ and $e_1$ and taking the limit $n\downarrow 0$, we get
\begin{subequations}
\begin{eqnarray}
e_1&=&\R_{\mt{R}}(-\chi_1)+\lambda \\
f_1&=&\sqrt{\gamma\sigma_u^2\R_{\mt{R}}(-\eta_1) + (q_1-\gamma\sigma_u^2\chi_1-\sigma_u^2p_1\mu_1)\R_{\mt{R}}^{\prime}(-\eta_1)}, \\
g_1&=&\sqrt{\frac{1}{\mu_1} \left[ \R_{\mt{R}}(-\chi)-\R_{\mt{R}}(-\eta_1) \right]}.
\end{eqnarray}
\end{subequations}
Following the same lines of derivation as in \cite{zaidel2012vector}, $\log \mathcal{M}(\mt Q)$ is straightforwardly derived as
\begin{eqnarray}
\log \mathcal{M}(\tilde{\mt{Q}})=\log \int \left[  \int \left( \sum_{x \in \mathbbmss{X}}\mathcal{K}(x,y,z) \right)^{\frac{\mu_1}{\beta}} \Dif y \right]^{\frac{n\beta}{\mu_1}} \Dif z
\end{eqnarray}
where the function $\mathcal{K}(x,y,z)$ is given by
\begin{eqnarray}
\mathcal{K}(x,y,z) \define \e^{2\beta \Re\{x(f_1z^*+g_1y^*)\} -\beta e_1 |x|^2}.
\end{eqnarray}
Consequently, by taking the derivative of the exponent term with respect to $f_1$, $g_1$ and $e_1$ the following fixed-point equations are concluded.
\begin{subequations}
\begin{eqnarray}
\chi_1+p_1\mu_1&=&\frac{1}{f_1}\int \int\Re\left\{z^*\argmin\limits_{x\in \mathbbmss{X}}|f_1z+g_1y-e_1x |   \right\}\tilde{\mathcal{Y}}(y,z)\Dif z \Dif y,\\
\chi_1+(q_1+p_1)\mu_1 &=& \frac{1}{g_1}\int \int\Re\left\{y^*\argmin\limits_{x\in \mathbbmss{X}}|f_1z+g_1y-e_1x |   \right\}\tilde{\mathcal{Y}}(y,z)\Dif z \Dif y,\\
q_1+p_1 &=& \int \int\left|\argmin\limits_{x\in \mathbbmss{X}}|f_1z+g_1y-e_1x |   \right|^2\tilde{\mathcal{Y}}(y,z)\Dif z \Dif y,\\
\int_{\chi_1}^{\eta_1} \R_{\mt{R}}(-w)\dif w&=&\int \log \int \mathcal{Y}(y,z)\Dif y \Dif z-2\chi_1\R_{\mt{R}}(-\chi_1)+ \lambda\mu_1(p_1+q_1) \nonumber \\
&&+(\mu_1q_1+2\chi_1+2\mu_1p_1-2p_1\mu^2\gamma\sigma_u^2-2\chi\mu\gamma\sigma_u^2)\R_{\mt{R}}(-\eta_1)\nonumber \\
&&-2\mu_1(q_1-\gamma\sigma_u^2\chi-\gamma\sigma_u^2p_1\mu_1)(\chi_1+\mu_1p_1)	\R_{\mt{R}}^{\prime}(-\eta_1),
\end{eqnarray}
\end{subequations}
where the function $\mathcal{Y}(y,z)$ is defined as
\begin{eqnarray}
\mathcal{Y}(y,z)\define\e^{-\mu_1 \min\limits_{x\in \mathbbmss{X}}e_1 |x|^2-2\Re\{x(f_1z^*+g_1y^*)\}}
\end{eqnarray}
and $\tilde{\mathcal{Y}}(y,z)$ denotes the normalized version of $\mathcal{Y}(y,z)$, i.e., 
\begin{eqnarray}
\tilde{\mathcal{Y}}(y,z)\define\frac{\mathcal{Y}(y,z)}{\int_{\mathbbmss{C}}  \mathcal{Y}(\tilde{y},z)\Dif \tilde{y}}.
\end{eqnarray}
By taking similar steps as in the RS-based analysis, the parameter $\tilde{D}$ is derived as
\begin{eqnarray}
\tilde{D}_{\rm RSB}&=& \gamma\sigma_u^2+\lim_{\beta\uparrow \infty} \lim_{n\downarrow 0}\frac{\alpha}{\beta} \frac{\partial}{\partial n}
n^2\beta^2f_1^2q_1+n\beta ( f_1^2p_1 \mu_1+ f_1^2\chi_1+ q_1 g_1^2\mu_1+ g_1^2 p_1\mu_1 + g_1^2)\chi_1\nonumber \\
&-& n\beta e_1(q_1+p_1) -n e_1 \chi_1 - \log \mathcal{M} (\tilde{\mt{Q}})+ n\lambda (\chi_1 +\beta (q_1+p_1))  +\int_0^{\gamma_1} \R_{\mt{R}}(-w)\dif w \nonumber \\
&+& \left( n-\frac{n\beta}{\mu_1}\right) \int_0^{\chi_1} \R_{\mt{R}}(-w)\dif w +
\left( \frac{n\beta}{\mu_1}-1\right) \int_0^{\chi_1+\mu_1p_1} \R_{\mt{R}}(-w)\dif w,
\end{eqnarray}
where we have used the subscript $\rm RSB$ to denote that $\tilde{D}$ is derived under the $1$-RSB assumption. After some lines of derivations, the asymptotic distortion under $1$-RSB is calculated as
\begin{eqnarray}
{D}_{\rm RSB}&=&\gamma\sigma_u^2-\frac{\alpha\chi_1}{\mu_1}
\R_{\mt{R}}(-\chi_1)+ \alpha\left(q_1+p_1+\frac{\chi_1}{\mu_1}-2\gamma\sigma_u^2 \eta_1 \right)\R_{\mt{R}}(-\eta_1) \nonumber\\
&& - \hspace*{.7mm} \alpha\eta_1(q-\gamma\sigma_u^2\eta_1)\R_{\mt{R}}^{\prime}(-\eta_1)
\end{eqnarray}
which concludes the proof.

\section{Lower-bound on the Asymptotic Distortion for $M$-PSK Signals}\label{appen3}
For the channel matrix $\mt H$ and the data vector $\matr u$, let $E_K$ denote the event that the minimum distortion over all possible $\matr{x} \in \mathbb{X}^N$ is less than $\epsilon$; then, one can write
\begin{eqnarray}
\Pr \left\lbrace E_K \right\rbrace =\Pr \left\lbrace \min\limits_{\matr{x}\in \mathbbmss{X}^N}\frac{1}{K}\|\mt{H}\matr{x}-\sqrt{\gamma}\matr{u}   \|^2\leq \epsilon\right\rbrace .
\end{eqnarray}
From the union bound, we have
\begin{eqnarray}
\Pr \left\lbrace E_K \right\rbrace = \Pr \left\lbrace \bigcup\limits_{\matr{x}\in \mathbbmss{X}^N} \frac{1}{K}\|\mt{H}\matr{x}-\sqrt{\gamma}\matr{u}   \|^2\leq \epsilon\right\rbrace \leq \sum\limits_{\matr{x}\in \mathbbmss{X}^N} \Pr \left\lbrace \frac{1}{K}\|\mt{H}\matr{x}-\sqrt{\gamma}\matr{u}   \|^2\leq \epsilon\right\rbrace.
\end{eqnarray}
For the $M$-PSK constellation, the entries of $\matr x$ lie on a circle in the complex plane. Therefore, for iid $\mt{H}$ and  Gaussian $\matr{u}$, the variable $\Delta=\|\mt{H}\matr{x}-\matr{u}   \|^2 / K$ in the large-system limit is a scaled chi-square random variable with $2K$ degrees of freedom. Assuming $\sigma_u^2=\gamma=1$, the probability density function of $\Delta$ reads
\begin{eqnarray}
f_{\Delta}(v)=\frac{K^K}{2^{K}\Gamma(K)}v^{K-1}\e^{-\frac{K}{2}v}.
\end{eqnarray}
Consequently, one can bound $\Pr \left\lbrace E_K \right\rbrace$ from above as
\begin{eqnarray}
\Pr \left\lbrace E_K \right\rbrace \leq M^{\alpha K}\int_{0}^\epsilon \frac{K^K}{2^{K}\Gamma(K)}v^{K-1}\e^{-\frac{K}{2}v} \dif v.
\end{eqnarray}
The function $v^{K-1}\e^{-\frac{K}{2}v}$ is an increasing function within some small neighborhoods of $v=0$. Therefore, a simple upper-bound for $\Pr \left\lbrace E_K \right\rbrace$ is given by
 \begin{eqnarray}
\Pr \left\lbrace E_K \right\rbrace\leq  M^{\alpha K}\epsilon \frac{K^K}{2^{K}\Gamma(K)}\epsilon^{K-1}\e^{-\frac{K}{2}\epsilon}.
 \end{eqnarray}
Asuming that $\alpha$ is bounded from above, one can the bound \cite{robbins1955remark}
\begin{eqnarray}
\Gamma(K)\geq \sqrt{2\pi}(K-1)^{K-1/2}\e^{-K+1},
\end{eqnarray}
and conclude that the argument
\begin{eqnarray}
\lim_{K\uparrow \infty} \Pr \left\lbrace E_K \right\rbrace = 0
\end{eqnarray}
holds, if we have
\begin{eqnarray}\label{alshar}
\frac{\epsilon}{2}-\log(\epsilon)>\alpha \log(M)-\log(2)+1
\end{eqnarray}
which concludes a lower-bound on the asymptotic distortion. In fact, for any $\alpha$ satisfying \eqref{alshar}, the distortion is larger than $\epsilon$ with probability one in the large-system limit. As the probability decays exponentially with $K$, it can be further shown that the statement holds also almost surely using the Barel-Cantelli lemma \cite{bremaud2012introduction}. Note that the bound is probably valid only for iid matrices.

\section{Channel with Path-loss Effect}\label{pathlosseff}
The effect of path-loss can be taken into account by considering $\mt H$ to be written as
\begin{eqnarray}
\mt{H} =\mt{D}^{1/2} \mt{U}, \label{pathloss_mat}
\end{eqnarray}
where $\mt{D}$ is a diagonal matrix whose $k$th diagonal entry $d_k$ is the normalized path-loss of the base station to $k$th user terminal, and $\mt{U}$ is a $K\times N$ iid matrix whose entries have variances equal to $1/N$. Let us assume that the users are uniformly located in an annular region around the base station. Denote the maximum and minimum distances to the base station with $r_{\max}$ and $r_{\min}$, respectively, and define the ratio of these distances to be $\kappa\define r_{\max} / r_{\min}$. Moreover, assume that the path-loss factors $d_1,\ldots,d_K$ are normalized such that the path-loss for the users located at the distance of $r_{\rm min}$ is $1$. Considering the path-loss exponent of $\nu$, one can derive the distribution of the path-loss factor $d_k$ for $k=1,\ldots,K$ as 
\begin{eqnarray}
f_{d}(d)=\frac{2}{\nu(\kappa^2-1)d^{2/\nu+1}}.
\end{eqnarray}
As the replica solutions only depend on the eigenvalues' distribution of the matrix $\mt{R}=\mt{H}^\dagger \mt{H}$, we need to calculate the $\rm R$-transform of $\mt{R}$ for $\mt H$ given in \eqref{pathloss_mat}. For this aim, we invoke Lemma~\ref{lemma3} from \cite{couillet2011random} to derive the Stieltjes transform of the eigenvalues' distribution of $\mt{R}$. 
\begin{lemma} \label{lemma3}
The Stieltjes transform of the matrix $\mt{R}=\mt{U}^\dagger \mt{D} \mt{U}$ satisfies
\begin{eqnarray}
\frac{1}{{\rm G}_{\mt{R}}(s)}+s= \alpha\int \frac{zf_d(z) {\rm d}z}{\alpha+z{\rm G}_{\mt{R}}(s)}.
\end{eqnarray}
\end{lemma}

Using Lemma~\ref{lemma3}, the $\rm R$-transform ${\R}_{\mt{R}}(w)$ is then numerically obtained from the Stieltjes transform using \eqref{Rstilet}. Using this strategy, the performance of the LSE precoders for the case with path-loss effect can be investigated. In this case, one can observe that the results with path-loss effect follow the same behavior as for the iid channel case.

\section{Generalization to Frequency-selective Fading Channels}{
\label{appenfreuencyselective}
Consider a case with frequency-flat fading channel. Let $L$ be the number of sub-carriers and assume that the fading at each frequency sub-band is frequency-flat. For simplicity, assume further that each sub-band includes one sub-carrier. Let $\mt{H}_j$ be the channel matrix at $j$th frequency sub-band whose entries are iid. The data input vector at the $\ell$th sub-carrier is denoted by $\matr{u}_\ell$. 

We consider a MIMO-OFDM approach in which the base station first precodes the data vectors for the sub-carriers, and then, uses one Inverse Fast Fourier Transform (IFFT) block per antenna. In this case, the LSE precoder needs to determine $L$ column vectors $\matr{v}_1,\cdots,\matr{v}_L$ to be given to the IFFT blocks as inputs. Denote the input vector for the IFFT block of the $\ell$th antenna by $\matr{v}_\ell$, and let $\matr{W}$ be the IFFT matrix. By defining $\matr{v}_{\mathrm{t}}\define{\mathrm{Vec}}\left([\matr{v}_1,\ldots,\matr{v}_L]^\T\right)$, 
the LSE precoder reads
\begin{eqnarray}\label{precfft}
\matr{v}_{\rm t}=\argmin\limits_{\matr{W}_{\rm t}\matr{x}_{\rm t}\in \mathbbmss{X}^{NL}} \|\mt{H}_{\rm t}\matr{x}_{\rm t}-\sqrt{\gamma} \hspace*{.5mm} \matr{u}_{\rm t}  \|^2 + \lambda\|\matr{x}_{\rm t} \|^2,
\end{eqnarray}
where $\mt{H}_{\rm t}$ is a $KL\times NL$ matrix whose $k$th part of its $(\ell-1)L+k$ columns is the $\ell$th column of the $\mt{H}_k$ and the remained entries are zero, 
$\matr{u}_{\rm t}\define[\matr{u}_1^\T,\ldots,\matr{u}_L^\T]^\T$ and $\matr{W}_{\rm t}$ is an $LN\times LN$ block-diagonal matrix whose $L\times L$ diagonal blocks are equal to $\matr{W}$. 
\eqref{precfft} can be rewritten
\begin{eqnarray}
\matr{v}_{\rm t}=\argmin\limits_{\matr{z}_{\rm t}\in \mathbbmss{X}^{NL}} \|\mt{H}_{\rm t}\matr{W}_{\rm t}^\dagger \matr{z}_{\rm t}-\sqrt{\gamma} \hspace*{.5mm} \matr{u}_{\rm t}  \|^2 +\lambda\|\matr{z}_{\rm t} \|^2, \label{eq:EqualFFad}
\end{eqnarray}
by using the fact that $\mt{W}_{\mathrm{t}}\mt{W}_{\mathrm{t}}^\dagger=\mt{I}$. Considering \eqref{eq:EqualFFad}, one can consider LSE precoding over an equivalent frequency-flat fading channel with the channel matrix equal to $\mt{H}_{\rm t}\matr{W}_{\rm t}^\dagger$. 

Fig.~\ref{eigen} compares the empirical cumulative distribution of the eigenvalues of $\mt{R}_{\rm t}=\mt{H}_{\rm t}^\dagger \mt{H}_{\rm t}$ and $\mt{R}_{j}=\mt{H}_{j}^\dagger \mt{H}_{j}$ numerically for $L=64$ and $K=N=128$ assuming that $\mt{H}_\ell$ for $\ell\in\{1,\ldots,L\}$ are iid matrices. It is observed that the eigenvalues' distribution in the both cases are the same for iid binary and Gaussian entries. As the results derived via the replica method depend only on the eigenvalues' distribution of $\mt{R}_{\rm t}$, the observation in Fig.~\ref{eigen} shows that the LSE precoder in this case has the same performance as in the case of frequency-flat fading channel. 

\begin{figure}[t]
\centering
\resizebox{.8\linewidth}{!}{
\pstool[width=0.6\linewidth]{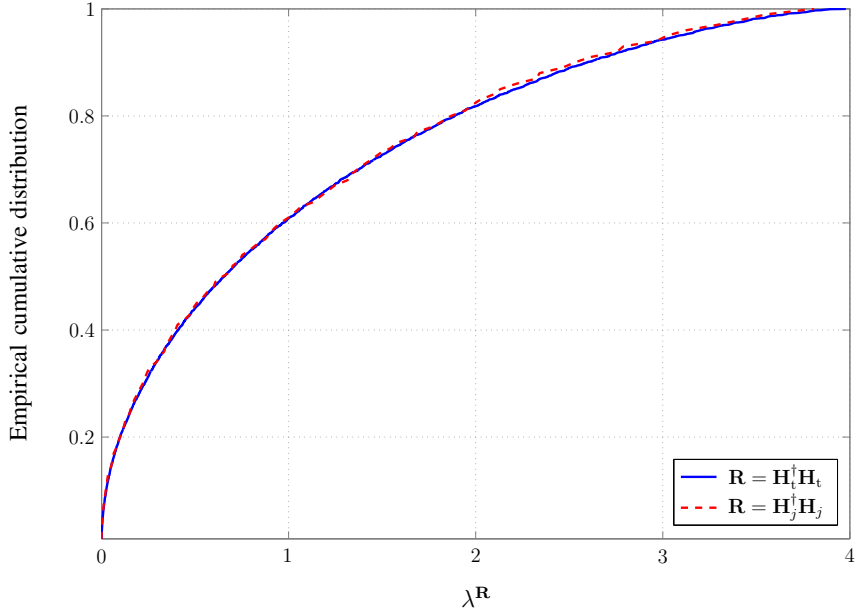}{
\psfrag{Empricial cumulative distribution}[c][c][.6]{Empirical cumulative distribution}
\psfrag{eigenvalues}[c][t][.6]{$\lambda^{\mt{R}}$}
\psfrag{AAAAAAA}[c][c][.5]{$\mt{R} =\mt{H}_{\mathrm{t}}^{\dagger}\mt{H}_{\mathrm{t}}$}
\psfrag{BBBBBBB}[c][c][.5]{$\mt{R} =\mt{H}_{j}^{\dagger}\mt{H}_{j}$}
\psfrag{0}[c][c][.5]{$0$}
\psfrag{1}[c][c][.5]{$1$}
\psfrag{2}[c][c][.5]{$2$}
\psfrag{3}[c][c][.5]{$3$}
\psfrag{4}[c][c][.5]{$4$}

\psfrag{0.2}[c][c][.5]{$0.2$}
\psfrag{0.4}[c][c][.5]{$0.4$}
\psfrag{0.6}[c][c][.5]{$0.6$}
\psfrag{0.8}[c][c][.5]{$0.8$}
\psfrag{1}[c][c][.5]{$1$}
}}
\caption{Empirical cumulative distribution of the eigenvalues of $\mt{H}_{\rm t}^{\dagger}\mt{H}_{\rm t}$ and $\mt{H}_{ j}^{\dagger}\mt{H}_{ j}$ for $L=64$ and $N=K=128$.}
\label{eigen}
\end{figure}
}
\end{appendices}

\bibliographystyle{IEEEtran} 
\bibliography{lit}

\end{document}